\documentclass[aps,prx,twocolumn,showpacs,amsmath,amssymb,superscriptaddress,longbibliography]{revtex4-2}

\usepackage{graphicx}

\usepackage{amssymb}
\usepackage{amsmath}
\usepackage{bm}
\usepackage{color}
\usepackage{xcolor}
\usepackage{multirow}

\newcommand{\uve}{\bm e}
\newcommand{\rb}{\bm r}
\newcommand{\vb}{\bm v}
\newcommand{\vm}{v_{\mathrm{max}}}

\newcommand{\Rsq}{R_{\rm sq}}
\newcommand{\kBT}{k_B T}
\newcommand{\Pe}{{\rm Pe}}
\newcommand{\OmA}{\Omega}

\graphicspath{{../figures/}}
\begin{document}
\title{Hydrodynamic pursuit by cognitive self-steering microswimmers}
\author{Segun Goh}
\email{s.goh@fz-juelich.de}
\affiliation{Theoretical Physics of Living Matter, Institute of Biological Information Processing and Institute for Advanced Simulation, Forschungszentrum J\"ulich, 
52425 J\"ulich, Germany}
\author{Roland G. Winkler}
\email{r.winkler@fz-juelich.de}
\affiliation{Theoretical Physics of Living Matter, Institute of Biological Information Processing and Institute for Advanced Simulation, Forschungszentrum J\"ulich, 
52425 J\"ulich, Germany}
\author{Gerhard Gompper}
\email{g.gompper@fz-juelich.de}
\affiliation{Theoretical Physics of Living Matter, Institute of Biological Information Processing and Institute for Advanced Simulation, Forschungszentrum J\"ulich, 
52425 J\"ulich, Germany}
\begin{abstract}
\noindent{\bf  ABSTRACT} \\[5pt]
The properties of biological microswimmers are to a large extent determined by fluid-mediated interactions, which govern their propulsion, 
perception of their surrounding, and the steering of their motion for feeding or in pursuit. Transferring similar functionalities to synthetic 
microswimmers poses major challenges, and the design of favorable steering and pursuit strategies is fundamental in such an endeavor. 
Here, we apply a squirmer model to investigate the pursuit of pursuer-target pairs with an implicit sensing mechanism and limited hydrodynamic 
steering abilities of the pursuer. Two hydrodynamic steering strategies are applied for the pursuer's propulsion direction by adaptation of 
its surface flow field, (i) reorientation toward the target with limited maneuverability, and (ii) alignment  with the target's propulsion 
direction combined with speed adaptation. Depending on the nature of the microswimmer propulsion (puller, pusher) and the velocity-adaptation scheme, 
stable cooperatively moving states can be achieved, characterized by specific squirmer arrangements and controllable trajectories. Importantly, pursuer 
and target mutually affect their motion and trajectories.
\end{abstract}
\maketitle 

\section*{Introduction} \label{sec:introduction}

The vast majority of motile biological microorganisms, such as bacteria, algae, or heterotrophic nanoflagellates, exploit fluid-mediated interactions for 
their propulsion, sensing of obstacles and prey, and feeding \cite{tuva:05,guas:10,kior:14}. Their micrometer size, 
which implies low-Reynolds-number fluid dynamics, where viscosity dominates over inertia, and strong thermal fluctuations \cite{elge:15,bech:16},  
renders an efficient hunting particularly difficult. An example is ``hydrodynamic starvation'' of  larval fishes, where hydrodynamic flow fields 
limit their feeding performance \cite{chin:14}. Another example is ambush-feeding, e.g., of copepods, which sense the hydrodynamic disturbances 
generated by the swimming prey and attack quickly \cite{kior:09}. Here, it is unclear how such microswimmers avoid warning the prey by hydrodynamic 
disturbances and pushing it away while attacking. Evidently, microscale predators are able to adjust their locomotion upon the sensed and gathered 
information to favorably approach the prey. Moreover, hydrodynamic interactions play a major role in the collective behavior of microswimmers 
\cite{thee:18,qi:22,sama:23}. This is paradigmatically reflected in  the behavior of bacteria with their large-scale swarming motion and active 
turbulence \cite{wens:12,dunk:13,qi:22,aran:22}. However, the role of signal exchange, aside from hydrodynamics, between microswimmers and of 
the corresponding adaption of motion and its effect on the emergent cooperative and collective behaviors is an issue which goes far beyond, and remains 
to be elucidated and understood.   

Similar questions arise in the design of synthetic intelligent micromachines (microbots), either via biomimetics  of biological microswimmers, e.g., flagellated swimmers \cite{rico:17,huan:22},  or by suitable colloidal realizations \cite{baeu:18,selm:18,lave:19,kasp:21}. Here, sensing and adaptation of motion is often achieved by stimuli via external fields \cite{baeu:18,selm:18}. Yet, paramount to perform complex tasks in biomedical and environmental applications is their ability to suitably adopt their motion to the changing surroundings along their trajectories \cite{erko:19,kurz:21}. This can be achieved by equipping the microbots either with engineered or biological actuators and sensors \cite{huan:22,cvet:14}. 

In the strive for the design of intelligent -- sensing and adaptive --  microswimmers~\cite{dai:16, Tsan:18,alva:21}, knowledge of favorable propulsion and steering mechanism is fundamental. A priori insight into emergent behaviors for a particular adaptation scheme would be beneficial for a desired performance, in particular, when hydrodynamic interactions and thermal noise play a major role.        

Here, we explore strategies and present simulation results for the pursuit by two types of microswimmers immersed 
in a fluid. We employ a generic model of microswimmers, squirmers \cite{ligh:52,blak:71,pak:14}, where the pursuer is equipped with implicit sensing and hydrodynamic response and adaptation mechanisms. Two different steering schemes are considered, extensions of the cognitive flocking model  -- active Brownian particles (ABPs), augmented by active reorientation of their directions of motion toward a moving object \cite{barb:16,goh:22} --, and of a hydrodynamic Vicsek model  \cite{vics:95,chat:20,chep:21} -- microswimmers which tend to align their swimming direction with their 
neighbors, augmented with speed adaptation.  For comparison, we also perform simulations of self-steering,
``intelligent" active Brownian particle (iABP) pairs, in particular, to resolve the effect of steric interactions 
on the microswimmer motion. Our simulations yield stable and unstable cooperative states, which strongly depend on the type of swimmer flow field (puller, pusher), and favorable choices of squirmer flow-field properties and steering suggest strategies for their stable motion and possibilities for target steering.

\section*{Results} \label{sec:results}

\subsection*{Model of self-steering microswimmer} \label{ssec:model}

\begin{figure}[t]
\centering
  \includegraphics[width= \columnwidth]{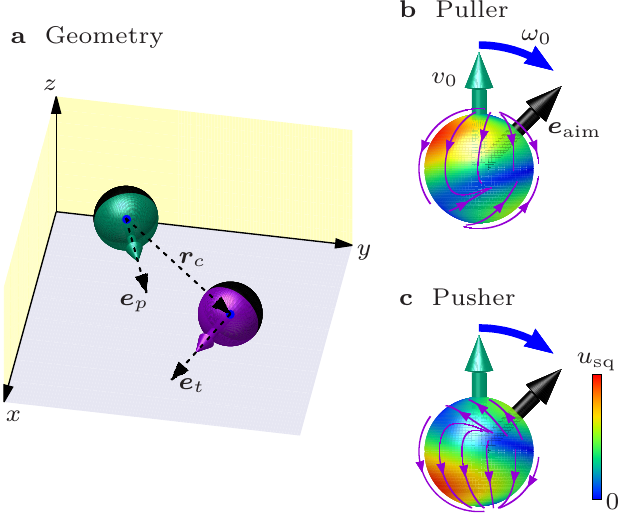}
  \caption{{\bf Illustration of the squirmers.}  
  {\bf a} The spherical pursuer (petrol) and target (purple) squirmer with the center-to-center difference vector ${\bm r}_c$ are self-propelled in the direction $\uve_p$ and $\uve_t$, respectively. 
  {\bf b}, {\bf c} Puller and pusher surface flow fields steering the propulsion direction $\uve_p$ in the direction $\uve_{\rm aim}$ via rotation around the vector $\boldsymbol{\omega}_0$ of frequency $\omega_0$. }
  \label{fig:sketch}
\end{figure}

A microswimmer is modelled as a spherical and neutrally buoyant squirmer \cite{ligh:52,blak:71,ishi:06,thee:18,shae:20}. The propulsion and steering are realized by an imposed surface flow field (slip velocity) \cite{pak:14}. Two types of 
particles are considered: pursuer and target. The target is propelled by an axisymmetric flow field along a body-fixed direction $\uve_t$ with the swim speed $v_t$, and moves, in absence of noise, along a straight trajectory.  Moreover, the surface flow field, described by the first two Legendre polynomials, creates an active stress, whose strength is characterized by the parameter $\beta$, where $\beta < 0$ corresponds to a pusher and $\beta >0$ to a puller (cf. Method Section). In contrast, the pursuer, with a body-fixed direction vector $\uve_p$ and swim speed $v_p$, is propelled by a non-axisymmetric flow field (Fig.~\ref{fig:sketch}). As a consequence, the pursuer moves on a helical trajectory in general. Pursuit of the target is achieved by adaptation of the pursuer's rotational motion via the non-axisymmetric flow modes, such that $\uve_p$ is redirected toward the desired direction, where the strength of the active reorientation is characterized by the maneuverability $\Omega$ (cf. Method Section). The embedding fluid is described by a particle-based mesoscale hydrodynamics simulation technique -- the multiparticle collision dynamics  approach --,  which 
captures the properties of fluctuating hydrodynamics \cite{kapr:08,gomp:09,huan:12,nogu:07,thee:16}.
Pursuer and target also experience hard-core repulsion to describe volume exclusion.

The goal of our study is, on the one hand, to provide insight into the emergent dynamics of the squirmer pursuer-target pair, 
and, on the other hand, to evaluate preferential pursuit strategies. We focus on three scenarios: (i) The pursuit of a target moving on a noisy ``straight'' trajectory. Here, the pursuer aims for the target and the speed ratio $\alpha = v_t/v_p$ is varied. (ii)  The pursuit of a target moving on a noisy ``helical'' trajectory, where  the helix slope $\xi = H/(2 \pi R)$ is varied, where $H$ is the helix pitch and $R$ the helix radius. (iii)  The pursuer aims to align its propulsion direction $\uve_p$ parallel to $\uve_t$ and adjusts its propulsion velocity to stay close to the target. The emergent dynamics of the pair is strongly affected by the interfering flow fields. In particular, the variation of the squirmer velocities is tightly connected with a change of their force-dipole strength, because $\beta$ is fixed, and a vanishing swimming velocity implies a vanishing active stress \cite{hu:15.1}.   

\subsection*{Noisy ``straight'' target trajectory} 
\label{ssec:straight}

\begin{figure*}[t]
\centering
  \includegraphics[width= \textwidth]{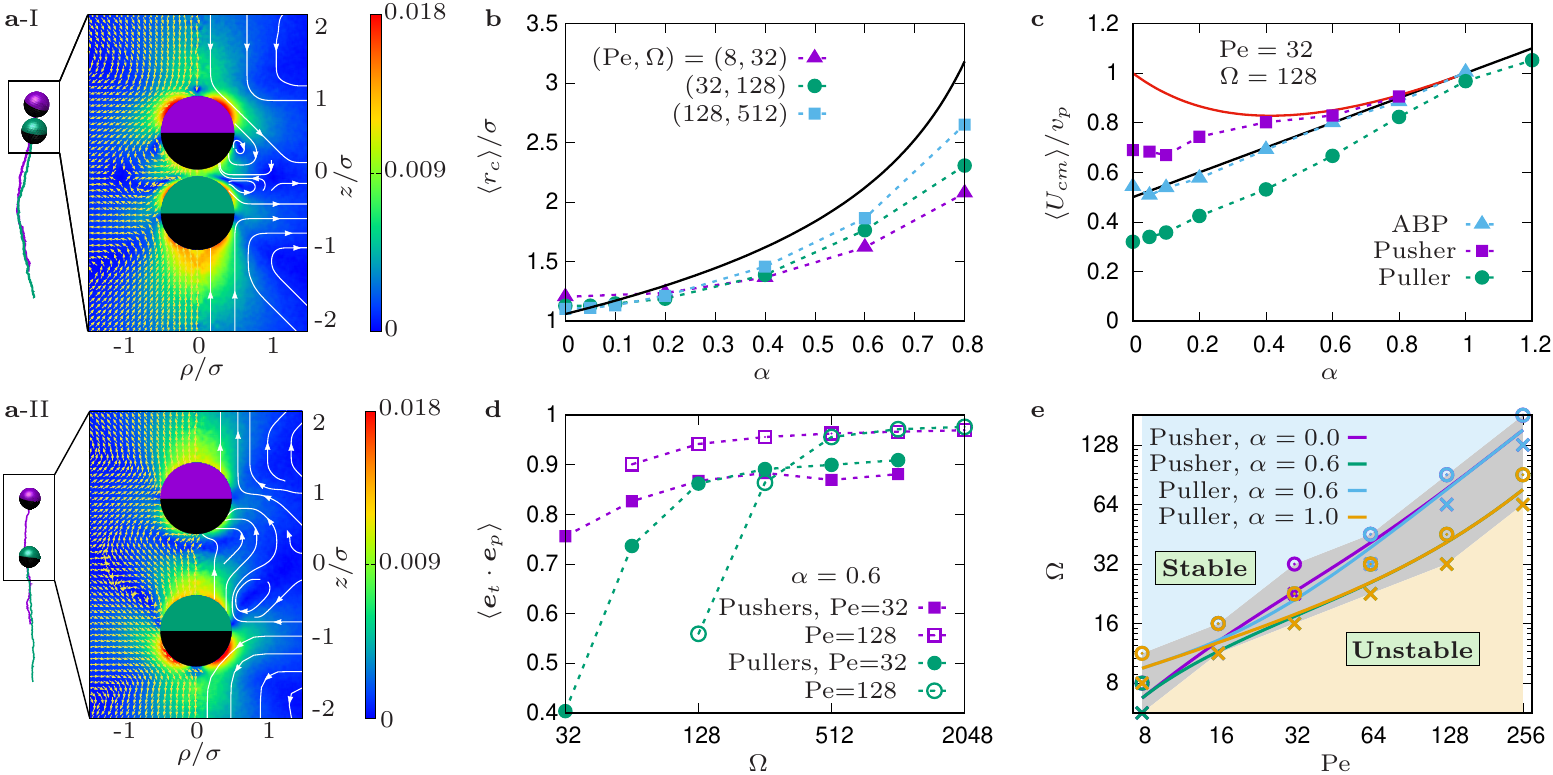} \\
  \caption{{\bf Target on a noisy ``straight'' trajectory.} 
  	{\bf a} Illustration of pursuer-target-pair trajectories and flow fields for {\bf I} pullers and {\bf II} pushers. 
  	{\bf b} Average pursuer-target distance $\langle r_c \rangle$ for pushers as a function of the speed ratio $\alpha = v_t/v_p$ for the indicated combinations of P\'eclet number and maneuverability (symbols). The black solid line represents the far-field approximation $r=r_0$ of Eq.~\eqref{eq:puller_distance}. 
  	{\bf c} Center-of-mass speed of  pursuer-target pairs for pushes (squares) and pullers (bullets). For comparison results of ABP pairs are shown (triangles). The black solid line indicates the speed $U_{cm}=v_p(1+\alpha)/2$ and the red solid line the far-field approximation of Eq.~\eqref{eq:mean_velocity}. 
  	{\bf d} Alignment parameter $\langle  \uve_p \cdot \uve_t \rangle$  as a function of the maneuverability $\Omega$ for pushers (squares) and pullers (bullets)  and the indicated P\'eclet numbers. No stable pairs are observed for $\Omega$ values smaller than $\Omega= 32$ for $\Pe =32$, $\Omega \approx 64$ (pushers) and $\Omega \approx 128$ (pullers)  for  $\Pe =128$. 
  	{\bf e} State diagram, which displays the boundary between regions of stable and unsuccessful pursuit as a function 
  	of $\Pe$ and $\Omega$, for pushers and pullers with various speed ratios $\alpha$, as indicated. 
	Pursuit is classified as unstable/unsuccessful when $r_c > 4\sigma$.
	Note that some symbols and lines lie on top of each other, in particular the data representing pusher pairs with $\alpha = 0.6$ and puller pairs with $\alpha = 1.0$ for $\Pe \geq 32$.
	}
  \label{fig:hydro_straight}
\end{figure*}
%Moreover, our finite-size periodic system limits the possible distances to $L_b/2 = 24 a = 3 \sigma$. 

The cooperative motion of pairs of non-steering squirmers has been analyzed in detail, both theoretically \cite{ishi:06,llop:10} 
and by simulations \cite{llop:10,goet:10,thee:16.1,clop:20}. In absence of thermal fluctuations, short-time stable attractive 
hydrodynamic interactions are predicted for parallel propulsion directions, when pullers move in a head-to-tail and pushers in a side-by-side configuration. However, the long-time asymptotic interaction between two squirmers is always repulsive, even though the transient regime can be rather long \cite{llop:10}.  Thermal fluctuations destabilize even the short-time cooperative motion, and the trajectories  of two nearby squirmers diverge rather quickly \cite{goet:10,thee:16.1,clop:20}. Similarly, 
suspensions of aligned self-propelled particles are always unstable to fluctuations, and a hydrodynamic instability due 
to particle active stresses is predicted for pushers but not for pullers \cite{sain:08}.
Hence, pursuit in absence of self-steering is essentially impossible, which would imply the extinction of any such ``dumb" predator.   

Self-steering of the pursuer toward the target changes the pursuit dynamics fundamentally, where the speed ratio $\alpha$ and the active stress play a decisive role for emerging stable cooperative states and a possible prey capture. For pullers ($\beta>0$), pursuer  steering toward the target implies a preferential head-to-tail cooperative state. This favors stable touching pursuer-target configurations due to their attractive hydrodynamic flow fields (Fig.~\ref{fig:hydro_straight}{\bf a-I}). Our simulations yield stable pairs even for speed ratios  $1 < \alpha \lesssim 1.4$ for $\beta=3$, i.e., for pursuer speeds smaller than the target speed (Fig.~S1, Movie 1). Here, self-steering dominates over both, (thermal) fluid fluctuations as well as destabilizing hydrodynamic torques \cite{ishi:06,llop:10}. For $\alpha \gtrsim 1.4$, no long-time stable configurations are found. This finding is qualitatively consistent with far-field hydrodynamic predictions with the assumption of a parallel alignment of $\uve_p$ and $\uve_t$ (SI, Sec.~S-IA3, which suggest stable stationary-state configurations for  
\begin{align} \label{eq:alpha_stable}
\alpha <  \frac{ (8/3)  (r_c/\sigma)^2 + \beta}{ (8/3)  (r_c/\sigma)^2  - \beta} ,
\end{align}
where $\sigma$ and $r_c$ denote the squirmer diameter and the instantaneous pursuer-target center-center distance, respectively (Fig.~\ref{fig:sketch}). For $\beta=3$ and reasonably close distance $r_c/\sigma = 2$, Eq.~\eqref{eq:alpha_stable} yields $\alpha< 1.78$, in semi-quantitative agreement with the simulation result (Fig.~\ref{fig:hydro_straight}{\bf b}). In addition to the obviously strong distance dependence, Eq.~\eqref{eq:alpha_stable} also neglects noise, which reduces the hydrodynamic interactions by disturbing the head-to-tail configuration. 

More challenging is the pursuit in case of pusher pairs ($\beta <0$), because a head-to-tail configuration for the maneuverability  
$\OmA =0$ is hydrodynamically unstable and squirmers repel each other. Yet, according to Eq.~\eqref{eq:alpha_stable}, stable pursuit 
is possible as long as $\alpha \lesssim 0.56$ for $\beta =-3$ and  $r_c/\sigma = 2$, 
which roughly corresponds to the average distances between two pusher squirmers obtained in the simulations (see Fig.~\ref{fig:hydro_straight}{\bf b}). 
Indeed, our simulations confirm successful pursuit even for $\alpha \lesssim 0.8$ (see Fig.~\ref{fig:hydro_straight}{\bf a-II} and Movie 2).  In the 
stationary state, the pursuer and target assume a finite average distance, as displayed in Fig.~\ref{fig:hydro_straight}{\bf b}, and the 
pursuer is not able to catch up with the target, but steering implies a stable cooperative state in a leader-follower arrangement. However, 
the pursuer hydrodynamically pushes the target, similarly to starving fish larvae, because its velocity is larger than that of the target. 
The average distance increases with increasing $\alpha$, i.e., when the pursuer speed is reduced. Note that $\alpha =0$ essentially corresponds 
to a resting target with the pursuer ``circling''  around it and touching it for large velocities $v_p \to \infty$, or P{\'e}clet numbers 
(Eq.~\eqref{eq:peclet}). In the noise-free far-field limit, Eq.~\eqref{eq:alpha_stable} yields the stationary-state pursuer-target distance 
(see SI, Eq.~(S27)) 
\begin{align} \label{eq:puller_distance}
r_c <  r_0 = \sigma \sqrt{\frac{3 \beta (\alpha +1)}{8 (\alpha -1)} } ,
\end{align} 
which qualitatively confirms our numerical result ($\beta=-3$), as displayed in Fig.~\ref{fig:hydro_straight}{\bf b},
where a pursuer has to be faster than the target, i.e., $\alpha < 1$, for stable pursuit.  

The effect of this interference of the squirmers' hydrodynamic flow fields on the center-of-mass speed $U_{cm}$ of 
the squirmer pair 
is displayed in Fig.~\ref{fig:hydro_straight}{\bf c}. In absence of active stress, the center-of-mass speed of the two squirmers would 
be $U_{cm}=(v_p +v_t)/2 =v_p(1+\alpha)/2$, which increases linearly with $\alpha$, in agreement with touching ABPs. 
Interference of the hydrodynamic flow fields leads to an enhancement of $U_{cm}$ for pullers and reduction for pullers. The speed
difference between pullers and pushers decreases with increasing $\alpha$, and for $\alpha \gtrsim 0.8$ all swimmer types 
move with the same center-of-mass velocity. We like to emphasize that pushers are well separated for $\alpha \gtrsim  0.3$ 
(see Figs.~\ref{fig:hydro_straight}{\bf b} and S1{\bf c}), and thus the target speed is enhanced via the fluid. It is this enhancement, 
which implies $U_{cm}$ to be larger than for ABP pairs.  With the assumption of pursuer-target arrangements with
$\uve_t \! \parallel \! \uve_p \! \parallel  \! \rb_c$, the hydrodynamic far-field approximation yields the center-of-mass 
speed (see SI Eq.~(S29))
\begin{align} \label{eq:mean_velocity}
\bar v = \frac{1+\alpha^2}{1+\alpha} v_p ,
\end{align}
which describes the simulation results very well for $\alpha > 0.6$ (see Fig.~\ref{fig:hydro_straight}{\bf c}). 
For $\alpha < 0.6$, $\uve_t$ is neither well aligned with $\uve_p$ nor with $\rb_c$ (Fig.~\ref{fig:hydro_straight}{\bf e}), 
which violates the premises of the theoretical result and explains the deviations from $\bar v$. The dynamics of puller pairs is 
dominated by near-field hydrodynamic effects, with attractive flow fields, which leads to $U_{cm}$ to be smaller than for ABP pairs.

A crucial parameter in self-steering pursuit is the maneuverability $\Omega$. By studying the dependence of the 
average distance $\langle r_c \rangle$ and the alignment parameter 
$\langle  \uve_p \cdot \uve_t \rangle$, on the P{\'e}clet number and $\Omega$, we can construct a phase diagram, which 
indicates the regions of stable and of unsuccessful pursuit. 
The dependence of the alignment of the propulsion directions $\uve_t$ and $\uve_p$ on the maneuverability is 
illustrated in Fig.~\ref{fig:hydro_straight}{\bf d} for the speed ratio $\alpha=0.6$ and various P\'eclet numbers. 
For sufficiently large $\Omega$, both pullers and pushers exhibit the same alignment parameter 
$\langle  \uve_p \cdot \uve_t \rangle$, nearly independent of $\Pe$, only slightly larger for larger $\Pe$. 
Noteworthy is the decrease of alignment with decreasing $\Omega$ for pullers, where the drop appears at 
large $\Omega$ for larger $\Pe$. Hence, $\Omega$ has to exceed a threshold value for a sufficiently strong 
alignment of the propulsion directions and the formation of stationary pairs.
This is reflected in the phase diagram of Fig.~\ref{fig:hydro_straight}{\bf e}.
Only for large maneuverability the destabilizing effects of thermal fluctuations and hydrodynamic torques can be
overcome and stable head-to-tail configuration can be achieved especially for pullers. Furthermore, the
phase diagram shows that
(i) pursuit is more difficult for smaller $\alpha$ (slower target), because the pursuer overshoots and circumvents 
slow targets with a long detour;
(ii) pushers display a better pursuit performance than pullers for the same $\alpha = 0.6$, 
because near-field hydrodynamic effects hinder pullers with small maneuverability from forming head-to-tail configurations 
(see SI, Fig. S5{\bf b-V}).

\begin{figure*}[t]
\centering
\includegraphics[width= \textwidth]{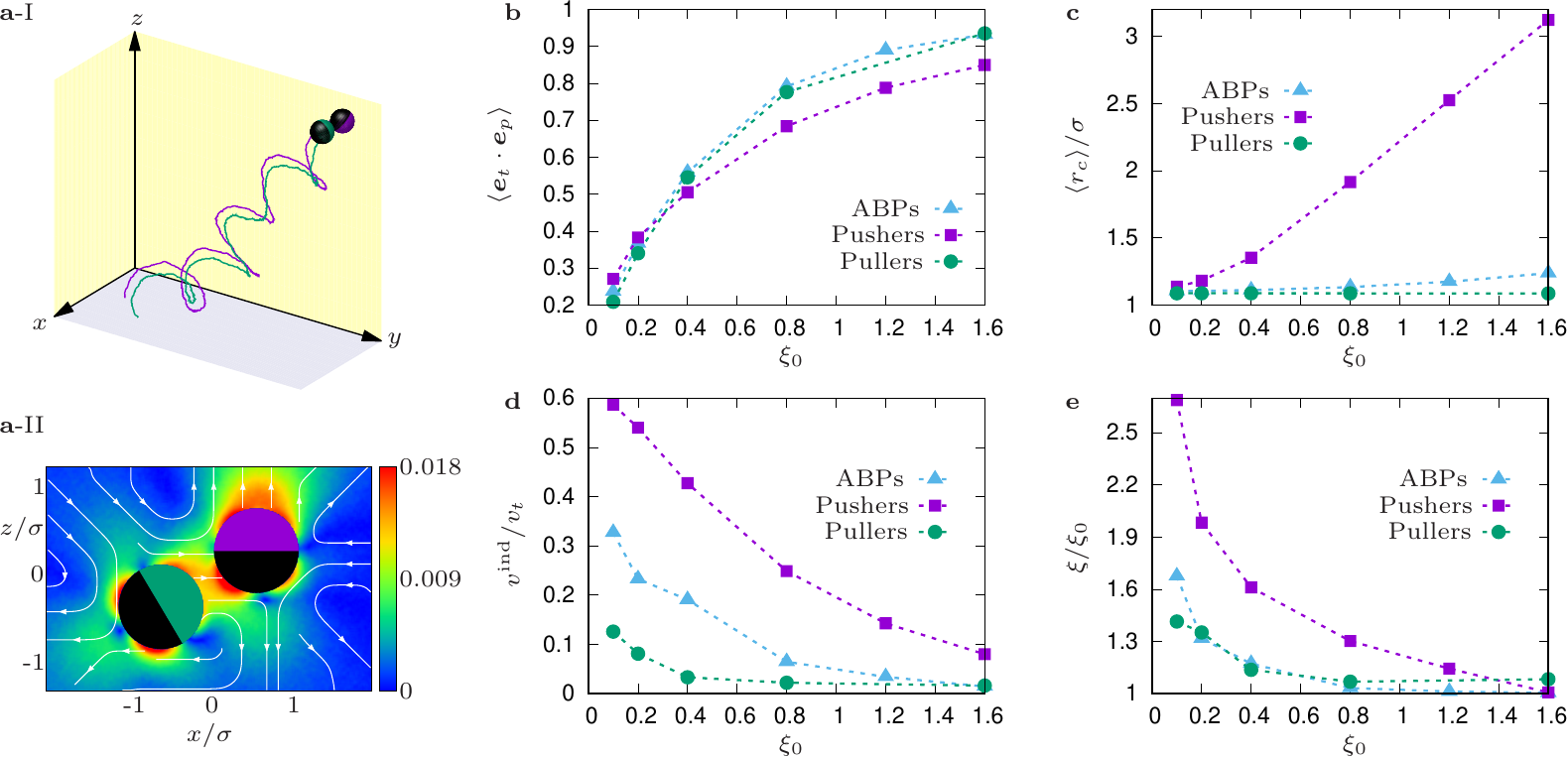}
\caption{{\bf Pursuer and target on helical target trajectory.}  
{\bf a-I} Illustration of pursuer-target-pair trajectories and {\bf a-II} their flow field. 
{\bf b} Order parameter $\langle  \uve_p \cdot \uve_t \rangle$, 
{\bf c} mean pursuer-target distance, 
{\bf d} hydrodynamic flow-field induced target velocity $v^{\rm ind}$, and 
{\bf e} effective helix slope $\xi$ as a function of the set slope $\xi_0 = H_0/(2\pi R_0)$ for pusher (squares), puller (bullets), 
and ABP (triangles) pairs. The parameters are ${\rm Pe} =128$, $\Omega = 8\,\Omega^{\rm hel}$, where 
$\Omega^{\rm hel} = {\rm Pe}/\sqrt{1+\xi_0^2}$ is the target angular velocity in units of $D_R$ 
(Eqs.~\eqref{eq:Omega_0}, \eqref{eq:hel_C1}, and \eqref{eq:hel_C2}), $R_0 = \sigma$, and $\alpha =1$.}
  \label{fig:helical_trajectory}
\end{figure*}

\subsection*{Noisy helical target trajectory} \label{ssec:helix}

By the nature of their flagella arrangement and the resulting chirality, bacteria often swim along helical trajectories rather than straight ones.  For example, the flagella of {\em E. coli} bacteria form bundles, which can be inclined with the cell body and, hence, leads to a wobbling motion along a helical trajectory \cite{turn:00,bian:17,math:19,mous:20,clop:21}. There are many more chiral microswimmers in nature with correspondingly helical trajectories \cite{cort:21,sama:23}. In such a situation, 
the pursuer and target do not have to move in a head-to-tail configuration, but the pursuer may follow a more favorable trajectory.
 
Figure~\ref{fig:helical_trajectory}{\bf a} shows an example of the trajectories of a pursuer and a target and their flow field, where the 
latter moves on a helical trajectory of radius $R$ and pitch $H$. Evidently, the pursuer traces the target on a helix with a radius, which is smaller than that of the target. Consequently,  it traverses a shorter trajectory than the target, which allows the pursuer to follow the target even for speed ratios $\alpha >1$. The particular arrangement is reflected in the relative orientation of the pursuer and target propulsion direction in Fig.~\ref{fig:helical_trajectory}{\bf b}.  For small helix slopes $\xi_0 = H_0/(2\pi R_0)$, $R_0$ and $H_0$ are the radius and pitch of an individual noise-free target, $\uve_p$ and $\uve_t$ are almost orthogonal independent of the P\'eclet number  (cf. Fig. S4{\bf b}), but the propulsion directions become increasingly parallel with increasing $\xi_0$, which corresponds to a head-to-tail configuration. This applies to all swimmer types, pushers, pullers, and ABPs. 
In the limit $\xi_0 \to 0$ and small radii $R_0 \gtrsim \sigma$, the target moves along a circle with the pursuer located in its 
center and the propulsion direction is essentially pointing in the radial direction.  The puller and ABP pursuer-target pairs touch each other (Fig.~\ref{fig:helical_trajectory}{\bf c}, see also Movie 3 for pullers), and steric interactions are important, which also affect the flow field of puller pairs, with an emerging Stokeslet flow field \cite{clop:20}. Even for pushers, this arrangement is particularly stable, because it reduces the pusher-pusher hydrodynamic repulsion. For large radii, $R_0 \gg \sigma$, the pursuer follows the target on a circle in a more head-to-tail configuration, as discussed above.
With increasing $\xi_0$, the target trajectory  straightens out. Pullers still attract each other and remain in a closely touching configuration for all $\xi_0$ (Fig.~\ref{fig:helical_trajectory}{\bf c}). The average ABP pursuer distance increases slightly due to thermal fluctuations. Pronounced flow-field effects emerge for pushers, with a pursuer-target distance larger than their diameter. The increasing repulsive interactions for the appearing head-to-tail configurations implies a substantial, almost linear growth of $r_c$ for $\xi_0 > 0.4$. As for straight trajectories, hydrodynamic interactions push the target, which renders a close approach increasingly difficult with increasing helical slope, but the pusher pair swims stable in a cooperative manner (see Movie 4).                     

The squirmer configuration also affects the helical trajectory of the target. Especially, a target velocity along the distance vector $\rb_c$, $v^\mathrm{ind} =\langle \vb_t \cdot \uve_c\rangle$, is induced, which is largest for pushers. This additional force on the target affects the properties of its trajectory. Although the helix radius is hardly affected for $\alpha =1$, the pitch increases substantially  for $\xi_0 \lesssim 0.5$, particularly for pushers.

The properties of the pursuer-target pair and the target trajectory also depend on the speed ratio $\alpha=v_t/v_p$ (see SI, Fig.~S4{\bf a} for $\xi_0=0.4$), with a qualitatively similar dependence as in Fig.~\ref{fig:helical_trajectory}, with $\alpha$ 
instead of $\xi_0$. 
For any kind of squirmer pair, the pursuer traverses a shorter path than the target, which allows it to follow the target even for $\alpha >1$. The induced velocity $v^\mathrm{ind}$ decreases with increasing $\alpha$ due to the weaker steric and hydrodynamic effects, in particular for $\alpha >1$, where the pursuer propulsion speed is smaller than that of the target (see Fig. S4{\bf a-III}).

\begin{figure*}[t]
\centering
  \includegraphics[width= \textwidth]{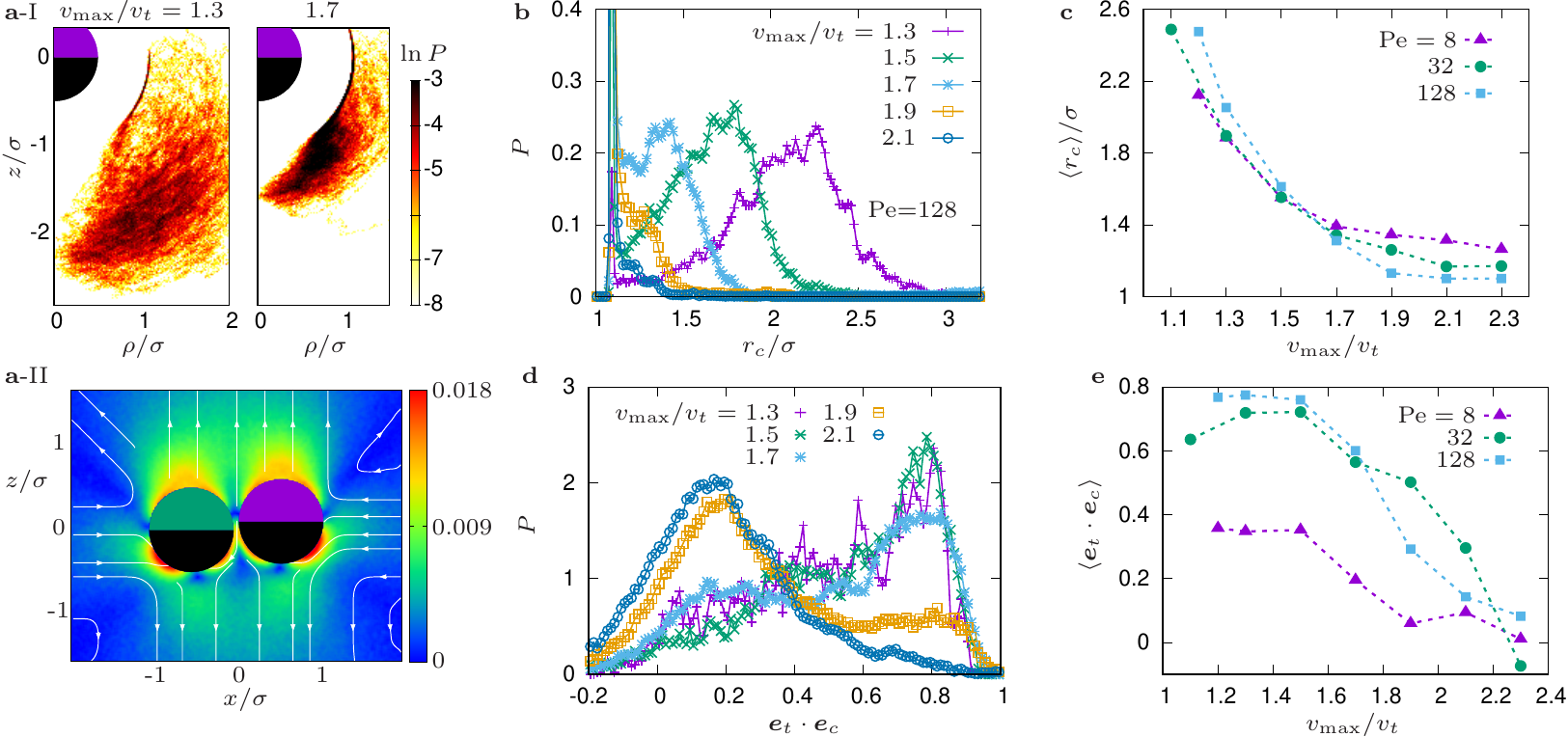}
 \caption{{\bf Pusher cooperative swimming by propulsion-direction alignment and speed adaption.} 
  {\bf a-I} Probability distribution of the pursuer position in cylindrical coordinates. The target is located at the origin of the reference frame. {\bf a-II} flow field of the pusher pair. 
  {\bf b} Probability distribution of the pursuer-target distance for the indicated speed ratios, where $\vm$ is the maximum of the adjustable pursuer speed. 
  {\bf c} Mean pursuer-target distance as a function of the maximum speed $\vm$ and various P\'eclet numbers. 
  {\bf d} Probability distribution of  $\uve_t \cdot \uve_c$ for various $\vm$.
  {\bf e} Order parameter $\langle \uve_t \cdot \uve_c \rangle$ as function of $\vm$ for various P\'eclet numbers. 
  Here, $\OmA =75.9$ and $\kappa = 1$ in all cases, and ${\rm Pe} = 128$ in {\bf a}, {\bf b} and {\bf d}.
  }
  \label{fig:alignment}
\end{figure*}

\subsection*{Propulsion direction alignment and speed adaptation} \label{ssec:align}

\begin{figure*}[t]
\centering
  \includegraphics[width= \textwidth]{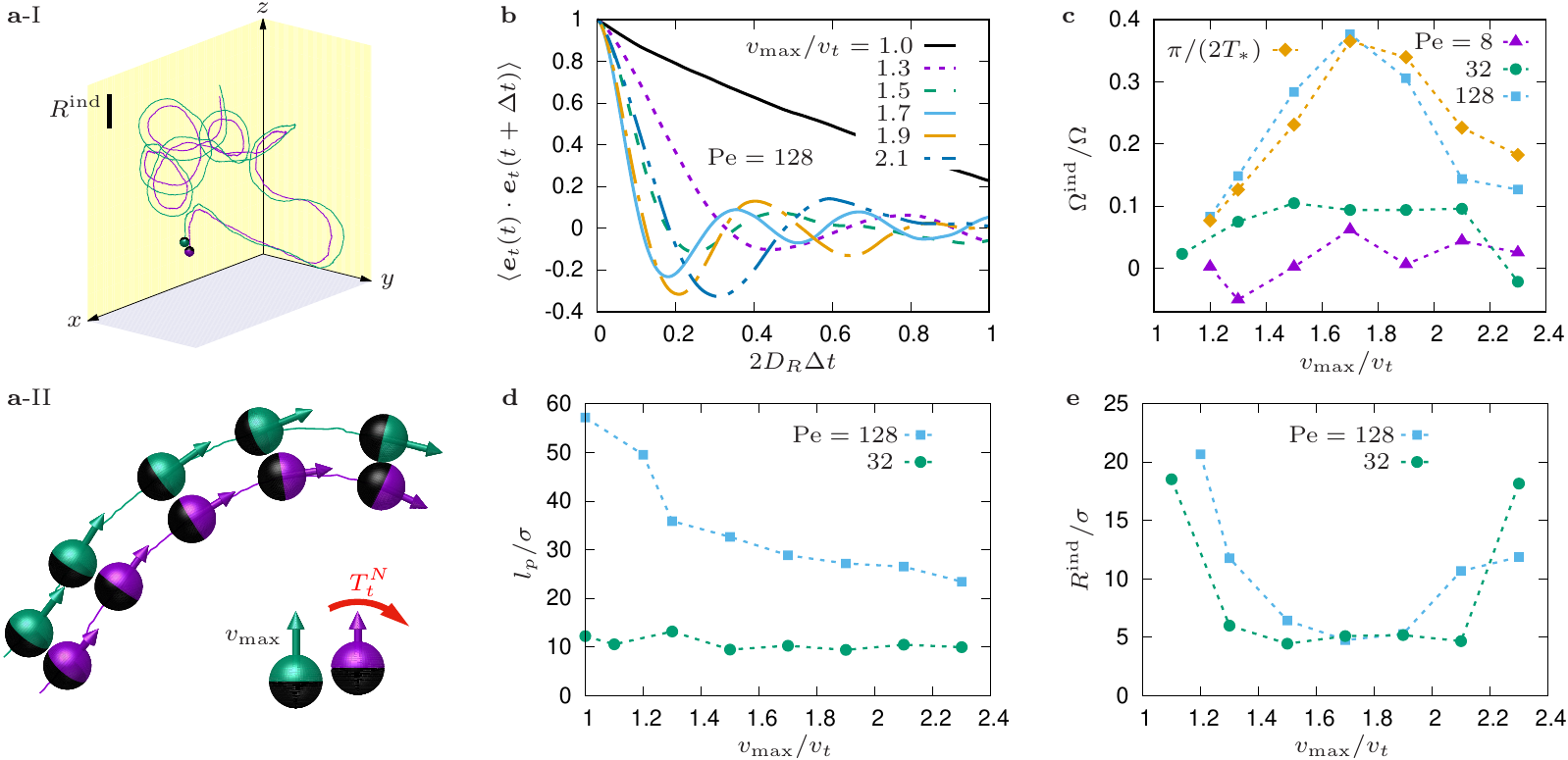}
  \caption{{\bf Pusher cooperative swimming by propulsion-direction alignment and speed adaption.} 
  	{\bf a} Emergence of cooperative circular motion. {\bf a-I} Example trajectory, where the vertical black bar represents the radius of induced circular motion (Fig.~\ref{fig:align}{\bf e}). {\bf a-II} Illustration of curvilinear trajectories 
  	induced by the torque $T_t^N$ (red arrow, bottom-right) of the pursuer (petrol) exerted on the target with $v_{\rm max}$ (purple). The pursuer is behind the target. 
  	{\bf b} Temporal autocorrelation function of the target propulsion direction $\uve_t$ as a function of the lag-time 
  	$\Delta t$, for ${\rm Pe}=128$ and various $\vm$, where $D_R$ is the rotational diffusion coefficient of an individual squirmer. 
  	{\bf c} Induced rotational frequency $\Omega_t^{\rm ind}$, obtained from Eq.~\eqref{eq:omega_ind} as a function of $\vm$ for various P\'eclet numbers as indicated (purple, green, and blue symbols),as well as extracted from the velocity autocorrelation functions in {\bf b} (dark yellow symbols). The first zero ($T_*$) of the velocity autocorrelation function is employed to determine the period. 
  	{\bf d} Persistence length $l_p$ obtained from the characteristic short-time decay of the velocity autocorrelation function. 
  	Here, an exponential function $\exp{(-\Delta t/\tau)}$ multiplied by a cosine function is fitted to the data at short times ($\langle {\bm e}_t (t)\cdot {\bm e}_t (t+\Delta t) \rangle < 1/{\rm e}$), from which $l_p$ is obtained as $l_p = v_t \tau$. 
  	{\bf e} Radii of induced circular motion determined from $R^{\rm ind} = v_t/\Omega^{\rm ind}$. 
  	In all cases, $\Omega = 75.9$ and $\kappa = 1$.}
  \label{fig:align}
\end{figure*}

The collective motion of flocks of birds or school of fish is governed by various interactions. In particular, the alignment of the moving direction of an individual with its neighbors plays a fundamental role \cite{ball:08.1,cava:14,papa:22,gomp:20,katz:11}. Theoretically, it has been shown in the Vicsek model \cite{vics:95,vics:12} and its extensions \cite{tone:95,chat:20,shae:20,chep:21} that such an interaction can lead to a preferred swimming direction with broken rotational symmetry. Particle-based alignment models typically neglected hydrodynamic interactions \cite{chat:20,chep:21}, which, however, are essential for the emergent collective behavior of microswimmers \cite{thee:18,qi:22}. Here, we address the influence of the microswimmer flow field on the cooperative motion of self-aligning squirmers and propose an adaptation scheme for stable motion. 

As the cooperative motion is always unstable for initially aligned squirmers \cite{ishi:06,llop:10,goet:10,thee:16.1,simh:02,sain:08} 
as well as for self-steering, persistently aligned squirmers (see SI, Sec.~S-VB1 for more details), we introduce a speed-adaption 
mechanism for the pursuer to achieve stable pursuit. Applying a velocity relaxation mechanism (see Methods, 
Eqs.~\eqref{eq:velosity_acceleration}, \eqref{eq:velocity}),  the propulsion speed $v_p(t)$ of the pursuer is changed by  acceleration/decelerated 
it, depending on its position with respect to the target expressed by the scalar product $\uve_p \cdot \uve_c$. As long as the pursuer is 
behind the target, $\uve_p \cdot \uve_c < 0$, it is accelerated, and while if it is in front, $\uve_p \cdot \uve_c > 0$, it is slowed down. 
An example for the probability of the pursuer position is presented in Fig.~\ref{fig:hydro_straight}{\bf a-I} and a trajectory in 
Fig.~\ref{fig:align}. Note that $\uve_p$  and $\uve_t$ are nearly parallel for the considered maneuverability. The propulsion velocity is 
limited by a maximum value $v_{\rm max}$ (see Methods, Eq.~\eqref{eq:velosity_acceleration}).      

A stable cooperative state requires an effective attraction between the microswimmers. Thus, no stable state is possible even for aligned 
propulsion directions of hard-sphere-like ABP particles. Pullers would require a head-to-tail arrangement for stable motion, which, however, 
already turns out to be unstable for persistently aligned squirmers via self-steering (see SI Sec.~S-VB1). Likewise, the proposed alignment 
scheme fails to yield a stable cooperative motion for pushers. 
Such a stable cooperative state is achieved for pushers by the proposed speed adaptation, where their relative arrangement depends on the 
maximum speed $v_{\rm max}$. Figure~\ref{fig:alignment}{\bf a} presents the probability distribution function of the pursuer position in 
cylindrical coordinates, and the pursuer-target flow field.  For a speed ratio $v_{\rm max}/v_t >1$, the pursuer is able to follow the 
target despite the presence of noise. The distance $r_c$ strongly depends on $v_{\rm max}$ and the probability distribution is broad -- the 
pursuer is even touching the target --, as shown in Fig.~\ref{fig:alignment}{\bf b}. The maximum of the distribution and the average 
distance decreases with increasing $v_{\rm max}$, depending only weakly on the P\'eclet number (Fig.~\ref{fig:alignment}{\bf c}). This is 
related to a change of the relative arrangement of the squirmers. For smaller $v_{\rm max}$ and larger $r_c$, the squirmers preferentially 
assume a head-to-tail configuration, as expressed by the distribution function of the  $\uve_t \cdot \uve_c$ and the corresponding order 
parameter  $\langle \uve_t \cdot \uve_c \rangle$  in Fig.~\ref{fig:alignment}{\bf d}, {\bf e}, see also Movie 5). However, the distribution function 
$P(\uve_t \cdot \uve_c)$ exhibits  a long tail toward small values of $\uve_t \cdot \uve_c$. Already for a small enhanced pursuer speed, 
$v_{\rm max}/v_t \approx 1.2$, the pursuer is able to surpass the target, which implies a transition between pursuer-target repulsion and 
attraction (Fig.~\ref{fig:alignment}{\bf a-I}). It is this dynamics, which leads to a stationary cooperative motion. With increasing  $v_{\rm max}$, 
the distance $\langle r_c \rangle$ and the average $\langle \uve_t \cdot \uve_c \rangle$ decrease. At the same time, the maximum of the 
distribution function $P(\uve_t \cdot \uve_c)$ shift to smaller arguments, and the pursuer and target preferentially assume a side-by-side 
configuration, where the two squirmers touch (Fig.~\ref{fig:alignment}{\bf a-I}, {\bf b}) and attract each other.

Speed adaptation partially cancels far- and near-field hydrodynamic forces between pursuer and target. However, steering and speed adaptation 
of the pursuer still give rise to a hydrodynamic torque on the target. A pursuers moving behind the target ($\langle \uve_p \cdot \uve_c \rangle >0$), 
rotates the target propulsion direction toward the opposite side of the pursuer (Fig.~\ref{fig:align}{\bf a-II}), i.e., $\uve_t$ changes. The 
reverse effect on the pursuer is small, because steering nearly compensates the torque at large enough maneuverability. 
Figure~\ref{fig:align}{\bf b} displays the temporal autocorrelation function of the propulsion vector $\uve_t$, illustrating the strong 
correlations in its dynamics, with circular parts of the trajectory (Fig.~\ref{fig:align}{\bf a-I}, Movie 6). The period of oscillatory behavior 
of the correlation function first decreases with increasing $v_{\rm max}$, then increases again for larger values, which corresponds to 
an increase and subsequent decrease of the curvature of the trajectory with $v_{\rm max}$. This is also reflected in the ``induced''  
target angular velocity 
\begin{align} \label{eq:omega_ind}
	\Omega_t^{\rm ind} = \frac{1}{D_R} \left\langle {\bm \Omega}_t \cdot \frac{{\bm e}_p \times {\bm e}_t}{|{\bm e}_p \times {\bm e}_t|} \right\rangle,
\end{align}
which is essentially a near-field effect (see SI, Sec. S-VB2). Figure~\ref{fig:align}{\bf c} confirms the presence of maximum in the angular velocity 
for large P\'eclet numbers, and shows that the hydrodynamic effect can be as large as $38\%$ of the pursuer's maneuverability. Thus, the 
propulsion direction and trajectory of a microswimmer can be manipulated hydrodynamically by a pursuing pusher microswimmer. However, the 
hydrodynamic effects also render the dynamics more noisy by perturbing the target motion, which may obscure the circular motion, in 
particular for small ${\rm Pe}$ and small $r_c$. 
Quantitatively, this is captured by the persistence length, which we determine via 
the characteristic decay time of the velocity autocorrelation function  (Fig.~\ref{fig:align}{\bf d}), and the diameter of the circular 
motion due to the induced angular velocity (Fig.~\ref{fig:align}{\bf e}). For $\rm Pe = 32$, the persistence length is essentially 
independent of $v_{\rm max}$, whereas for $\rm Pe= 128$, $l_p$ decreases with increasing $v_{\rm max}$. This happens because the 
pursuer-target distance decreases with increasing $v_{\rm max}$ (Fig.~\ref{fig:alignment}{\bf c}), and larger ${\rm Pe}$ induce a larger effective noise.

A similar hydrodynamic effect has been found for two or more sedimenting colloidal particles or connected chains of colloids in a viscous fluid. This implies, for example, that sedimenting colloids with a small vertical displacement do not follow the direction of the gravitational field, but sediment with a sideways drift, while sedimenting semiflexible colloid chains can even display helical trajectories \cite{ekie:08,sagg:15}.

\section*{Discussion}

We have analysed the interplay of hydrodynamic propulsion, cognition and self-steering, 
limited maneuverability, hydrodynamic interactions, and thermal or active noise, for a
system of two microswimmers -- a moving target and a self-steering pursuer. 
Both microswimmers are modelled as squirmers, where the pursuer is able to adjust 
its hydrodynamic propulsion to follow the target. 
Implicitly sensed information about either the target position or the orientation of its 
propulsion direction is employed for steering of the pursuer through a non-axisymmetric 
modification of the surface slip velocity. The squirmer's hydrodynamic propulsion type 
(pusher or puller) together with the steering mechanism determines the emerging cooperative 
states. 

More specifically, for pursuers with active reorientation of its propulsion direction toward 
the target, we show that pullers are able to catch up with the target, which is advantageous 
for a predator to reach its prey. In contrast, for pushers the average pursuer-target 
distance is typically found to be finite, with the formation of stable cooperatively moving 
pairs -- and the only option for the pursuer to catch up with the target is for speeds much larger than the target speed.
This is related to the problem of starving fish larvae, which 
are typically pushers. 

For steering through alignment of the propulsion direction, the pursuit of a moving 
target is predicted to be always unstable for pullers, while it can lead to stable 
cooperative motion for pushers -- but only in combination with speed adaptation. Here, 
indirect steering of the target is possible via hydrodynamic torques exerted by the pursuer. 
This effect could be utilized for microbots for the guidance of active target particles 
toward a preselected location.

%We have analyzed the emergent behavior of a pair of microswimmer pursuer-target pair, where the pursuer is able to adjust its hydrodynamic propulsion via surface slip velocity to follow the target, applying the squirmer model. Implicitly sensed information about the target positions or the orientation of its propulsion direction leads to non-axisymmetric steering of the pursuer. The squirmers hydrodynamic flow field (pusher vs puller) determines the emerging cooperative state. Pullers with the propulsion direction oriented toward the target are able to catch, which is advantageous for a predator to reach its prey. However, the average distance between pushers is typically finite, with the formation of stable cooperatively moving pairs, and it is only able to catch up with the target for pursuer speeds much larger then that of the target. This would lead to extinction of the predator,  similarly to starving fish larvae. Propulsion-direction alignment (hydrodynamic Vicsek model) and speed adaptation leads to stable pusher cooperative motion, with indirect steering of the target via hydrodynamic torques exerted by the pursuer. This effect can be utilized for guiding an active particle in a non-invasive way. 
       
Hydrodynamic interactions  between the squirmers and thermal noise strongly influence their cooperative behavior, and is crucial for the emergent dynamics in systems of self-propelled microorganisms with information exchange. Our analysis of the various steering and adaptation schemes provides insight into useful pursuit strategies, in particular, reveals ineffective ones. This may help in the design of synthetic, information processing  microrobots at low Reynolds numbers, where hydrodynamic interactions and thermal fluctuations are  paramount.

\section*{Methods} \label{sec:method}

\subsection*{Intelligent squirmer} 
\label{sec:model_isq}

The dynamics of a squirmer is governed by its surface flow field ${\bm u}_{sq}$ \cite{ligh:52,blak:71,pak:14}. In case of a spherical, nonaxisymmetric squirmer of radius $R_{\rm sq}$, which translates and rotates, the spherical components of the surface  slip velocity in a body-fixed reference frame are (${\bm u}_{sq}= (u_\theta, u_\phi,0)^T$) \cite{pak:14}
\begin{align} \label{eq:axisymmetric}
%u_{{\rm sym},\theta}^b =& ), \\  \label{eq:squirming_velocity_th}
u_\theta =& \ B_1 \sin{\theta}\,(1+\beta \cos{\theta})	-\frac{1}{\Rsq^2}(\tilde{C}_{11}\cos{\phi} - C_{11}\sin{\phi}), \\ 
u_\phi =& \ \frac{\cos{\theta}}{\Rsq^2}(C_{11}\cos{\phi} +\tilde{C}_{11}\sin{\phi}) 
                 + \frac{\sin{\theta}}{\Rsq^2}C_{01}  , \label{eq:squirming_velocity_ph}  
\end{align}
where $\theta$ and and $\phi$ are the polar and azimuthal angles. $B_1=3 v_0/2$ determines the magnitude of the self-propulsion velocity $v_0$ and $\beta$ the stresslet, where $\beta > 0$ for pullers and $\beta < 0$ pushers. The non-axisymmetric components $C$ and $\tilde C$ are associated with 
the angular velocity 
\begin{align}
\boldsymbol{\omega}_0 &= \frac{1}{\Rsq^3}(C_{11}\uve_x +\tilde{C}_{11}\uve_y -C_{01}\uve), \label{eq:Omega_0}
\end{align}
of the squirmer body.
In a Cartesian laboratory reference frame, with an instantaneous squirmer motion along the $z$ direction, the parameters $C_{01}$, $C_{11}$, and $\tilde{C}_{11}$ determine the angular velocity along the unit vectors $\uve\equiv \uve_z$, $\uve_x$, and $\uve_y$ of the axis of the reference frame.

Steering of a squirmer, in the spirit of the cognitive flocking model \cite{barb:16,goh:22}, is achieved by a modification of the surface flow field. Biological microswimmers, such as {\em Chlamydomonas} and {\em Volvox} algae, stir the ambient fluid 
in the opposite side of the direction they intend to turn to. For an intelligent squirmer (iSquirmer) with limited maneuverability, the required angular velocity to turn its propulsion direction $\uve$ toward a desired direction ${\bm e}_{\rm aim}$ can be written as
\begin{align} \label{eq:Omega_aim}
\boldsymbol{\omega}_0 = C_0 {\bm e} \times {\bm e}_{\rm aim},
\end{align}
where $C_0$ quantifies the strength of the active reorientation, thus, determines the 
maneuverability.  
A comparable adaptation is achieved by  equating the non-axisymmetric components of surface flow field of Eq. \eqref{eq:Omega_0} with Eq.~\eqref{eq:Omega_aim}, which yields
\begin{align} \label{eq:Cs11}
C_{11} &= C_0 \Rsq^3 ({\bm e} \times {\bm e}_{\rm aim}) \cdot {\bm e}_x, \\ \label{eq:Cs11t}
\tilde{C}_{11} &= C_0 \Rsq^3 ({\bm e} \times {\bm e}_{\rm aim}) \cdot  {\bm e}_y , \\ \label{eq:Cs01}
C_{01} &= 0 .
\end{align}
Note that the rotation around the squirmer body axis, determined by $C_{01}$, is irrelevant,
because it does not affect the self-propulsion direction.

To quantify self-propulsion and self-steering of iSquirmers, 
we introduce the P{\'e}clet number 
\begin{align} \label{eq:peclet}
\Pe = \frac{v_0}{\sigma D_R},
\end{align}
and the maneuverability
\begin{align}
\OmA = \frac{C_0}{D_R} .
\end{align}
Here, $D_R$ denotes the rotational diffusion coefficient, which is determined by the fluid viscosity $\eta$ and the squirmer diameter $\sigma = 2R_{\rm sq}$.

In case of pursuit as in the cognitive flocking mode, the propulsion direction is oriented toward the target position, hence, 
\begin{align}
 {\bm e}_{\rm aim}= {\bm e}_{ c} = \frac{{\bm r}_c}{|{\bm r}_c|} , 
\end{align}
where ${\bm r}_c = {\bm r}_t-{\bm r}_p$  denotes the vector connecting the pursuer and target positions ${\bm r}_p$ and ${\bm r}_t$ (Fig.~\ref{fig:sketch}).

Velocity alignment, as in the Vicsek model \cite{vics:95}, is achieved by the condition\
\begin{align}
 {\bm e}_{\rm aim}= {\bm e}_t . 
\end{align}
Here, in the absence of both noise, the two active agents become perfectly aligned and self-propel along the common direction at the same velocity.

\subsubsection*{Squirmer on helical trajectory}
\label{app:helical}

A target moving on a helical trajectory of radius $R$, pitch $H$, and the slope  $\xi = H/(2\pi R)$ is achieved by the choice of its surface flow field: 
\begin{align}
&C_{{01}} = \frac{\Rsq^2}{2} \frac{\xi}{1+\xi^2} v_t, \quad
C_{{11}} = \frac{\Rsq^2}{2} \frac{1}{1+\xi^2} v_t, \label{eq:hel_C1} \\ 
&\tilde{C}_{11} = 0 . \label{eq:hel_C2}
\end{align}
The corresponding helix parameters are
\begin{align}
R &= \frac{\sqrt{(C_{{11}})^2 +(\tilde{C}_{{11}})^2}}{(C_{{01}})^2 + (C_{{11}})^2 +(\tilde{C}_{{11}})^2} v_t \Rsq^3, \\
H &= \frac{2\pi C_{{01}}}{(C_{{11}})^2 + (\tilde{C}_{{11}})^2 + (C_{{01}})^2} v_t \Rsq^3.
\end{align}
%%Here, we define
%%\begin{align} \label{eq:Omega_helical}
%%\Omega_t^{\rm hel} \equiv \frac{|\boldsymbol{\omega}_{0,t}|}{D_R} =  \frac{\rm Pe}{\sqrt{1+\xi^2}},
%%\end{align}
%%which is the magnitude of target angular velocity in dimensionless units.

\subsubsection*{Pursuer speed adaptation}

We model the speed adaptation ability of the  pursuer by a continuous-time acceleration/deceleration process: 
\begin{align} \label{eq:velosity_acceleration}
\dot{v}_{p} (t) = \left\{  \begin{array}{cc} 
\displaystyle 
\kappa \frac{v_{\rm max} - v_p(t)}{\Rsq} v_p (t), &\ \textrm{for}\ {\bm e}_p\cdot{\bm e}_c \geq 0, \\ 
\displaystyle 
-\kappa \frac{v_p (t)}{\Rsq} v_p (t), &\ \textrm{for} \ {\bm e}_p\cdot {\bm e}_c < 0 .
\end{array} \right.
\end{align}
The (dimensionless) friction coefficient $\kappa$ and the maximum speed $v_{\rm max}$ account for a limited 
speed-adaptation capacity.
With an initial condition $v_p (0) = v_p^0$, the above equations imply
\begin{equation}
v_p (t) = v_{\rm max} \left[ 1+ (v_{\rm max}/v_p^0 -1) \exp{(-v_{\rm max}(\kappa/\Rsq) t)} \right]^{-1}
\end{equation}
for persistent acceleration, and
\begin{equation}
v_p (t) = \left[ 1/v_p^0 + (\kappa /\Rsq )t \right]^{-1}
\end{equation}
for persistent deceleration, respectively.
Numerically, the speed $v_p(t)$ is obtained by the Euler scheme
\begin{align} \label{eq:velocity}
v_p (t+h) = v_p(t) +h \dot{v}_p (t) , 
\end{align}
with  the MPC collision time $h$. As initial condition, we choose $v_p(0) = v_t$. 

\subsubsection*{Steric squirmer repulsion}
\label{sec:steric}

Steric squirmer-squirmer  interactions are described by  the separation-shifted  Lennard-Jones potential
\begin{align} \label{eq:pot}
U_{\rm LJ}(d_s)=4\epsilon_0 \left[ \left( \frac{\sigma_0}{d_s + \sigma_0} \right)^{12} 
- \left( \frac{\sigma_0}{d_s + \sigma_0} \right)^6 +\frac{1}{4}\right],
\end{align}
for $d_s < (2^{1/6}-1) \sigma_0$ and zero otherwise. The distance $d_s$ between the  two closest points on the surfaces of two 
interacting spheres is chosen as  $d_s = r_c - \sigma - 2d_v$, where  $\sigma$ is the diameter of a squirmer, $r_c$ the 
center-to-center distance between the two squirmer, and $d_v$ a virtual safety distance to prevent loss of hydrodynamic 
interactions at close distances \cite{thee:16.1,thee:18,qi:20}. 

The solid-body equations of motion of the squirmers, their center-of-mass translational and their rotational motion described 
by quaternions, are solved by the velocity-Verlet algorithm \cite{thee:16.1,qi:20}.  

\subsection*{Fluid model: Multiparticle collision dynamics} \label{sec:mpc}

The fluid is modeled via the multiparticle collision dynamics (MPC) method, a particle-based  mesoscale simulation approach 
accounting for thermal fluctuations \cite{kapr:08,gomp:09}, which has  been shown to correctly capture hydrodynamic 
interactions \cite{huan:12}, specifically for active systems 
\cite{gold:09,reig:12,geye:13,brum:14,pak:14,thee:14,eise:16,hu:15,hu:15.1,mous:20,babu:12,rode:19,qi:22}.

We apply the  stochastic rotation variant of the MPC  approach with angular momentum  conservation (MPC-SRD+a) 
\cite{thee:16,nogu:08,qi:22}. The algorithm proceeds in two steps --- streaming and  collision. In the streaming step, 
the  MPC point particles of mass $m$ propagate ballistically over a time interval $h$, denoted as collision time. In the 
collision step, fluid particles are sorted into the cells of a cubic  lattice of lattice constant $a$ defining the 
collision environment, and their relative velocities, with respect to the center-of-mass velocity of the collision cell, 
are rotated around a randomly oriented axes by a fixed angle $\alpha$. The algorithm conserves mass, linear, and angular 
momentum on the collision-cell level, which implies hydrodynamics on large length and long time scales \cite{kapr:08,huan:12}. 
A random shift of the collision cell lattice is applied at every collision step to ensure Galilean invariance \cite{ihle:03}. 
Thermal fluctuations are intrinsic to the MPC method.  A cell-level canonical thermostat (Maxwell-Boltzmann scaling (MBS) 
thermostat) is applied after every collision step, which maintains the temperature at the desired value \cite{huan:10.1}. 
The MPC method is highly parallel and is efficiently implemented on a graphics processing unit (GPU) for a high-performance 
gain \cite{west:14}.  

Squirmer-fluid interactions appear during streaming and collision. While streaming squirmers and fluid particles, fluid particles are reflected at a squirmer's surface by applying the bounce-back rule and adding  the surface velocity $\bm u_{sq}$.  To minimize slip, phantom particles are added inside of the squirmers, which contribute when collision cells penetrate squirmers. In all cases, the total linear and angular momenta are included in the squirmer dynamics. More details are described in Ref.~\cite{thee:16.1} and the supplementary material of Ref.~\cite{qi:20}.

%We would like to emphasize that the translational velocity and the angular velocity of a squirmer in our simulations are not identical to $v_0 {\bm e}$ and $\boldsymbol{\omega}_0$, due to the presence of noise and steric interactions. We denote the translational and rotational velocities extracted from simulation as $\mathbf{U}$ and $\mathbf{W}$, respectively.??

\subsection*{Simulation setup and parameters}
\label{sec:parameter}

The average number of MPC particles in a collision cell is chosen as $\langle N_c \rangle = 50$, the collision time as $h= 0.02 a\sqrt{m /(k_B T)}$, and the rotation angle as $\alpha_c = 130^\circ$, where $a$ is the length of a cubic collision cell,  $T$ the temperature,  and $k_B$ the Boltzmann constant \cite{thee:18}. These  values yield the fluid viscosity of $\eta = 111.3 \sqrt{mk_B T}/a^2$~\cite{nogu:08,thee:15} (cf. SI for more details of the MPC algorithm). 
With the squirmer radius $\Rsq = 3 a$, the theoretical rotational diffusion coefficient is given as $D_R = 1.3 \times 10^{-5} \sqrt{\kBT/m}/a$, in close agreement with simulation results  \cite{thee:18}. We consider the P{\'e}clet numbers $\Pe = 8$, $32$, and $128$, which correspond to the  Reynolds numbers ${\rm Re} =0.0017$, $0.007$, and $0.027$, and the magnitude of the active stress $|\beta| = 3$. The length of the three-dimensional cubic simulation box is $L_b =  48 a = 16\Rsq$ periodic boundary conditions are applied. 

The parameter of the Lennard-Jones potential  \eqref{eq:pot} are set to $\sigma_0=0.5 a$, $\epsilon_0 =5 k_BT$, and $d_v=0.25 a$.

A passive sphere is neutrally buoyant with mass $M=5655 m$, where $m$ is the mass of a MPC particle, and the MPC time step $h$ is used in the integration of the squirmers' equations of motion.

%%%\section*{Acknowledgments}

%%%This work has been supported by the DFG priority program SPP 1726 ``Microswimmers -- from Single 
%%%Particle Motion to Collective Behaviour''. The authors gratefully acknowledge the computing time 
%%%granted through JARA-HPC on the supercomputer JURECA at Forschungszentrum J\"ulich.

\section*{Data availability}

The data that support the findings of this study are available from the corresponding author upon reasonable request.

\section*{Author contributions}
R.G.W. and G.G. designed the research. S.G. performed simulations and analysed the data.
All authors discussed results and wrote the manuscript together.

\section*{Competing interests}

The authors declare no competing interests.

\section*{Additional information}

Supplementary information is available for this paper. % at ????.

%\bibliographystyle{apsrev4-2}
%%\bibliography{/Users/winkler/ownCloud/publications_library/bibliography/bibliography}
%\bibliography{bibliography}

\begin{thebibliography}{77}%
	\makeatletter
	\providecommand \@ifxundefined [1]{%
		\@ifx{#1\undefined}
	}%
	\providecommand \@ifnum [1]{%
		\ifnum #1\expandafter \@firstoftwo
		\else \expandafter \@secondoftwo
		\fi
	}%
	\providecommand \@ifx [1]{%
		\ifx #1\expandafter \@firstoftwo
		\else \expandafter \@secondoftwo
		\fi
	}%
	\providecommand \natexlab [1]{#1}%
	\providecommand \enquote  [1]{``#1''}%
	\providecommand \bibnamefont  [1]{#1}%
	\providecommand \bibfnamefont [1]{#1}%
	\providecommand \citenamefont [1]{#1}%
	\providecommand \href@noop [0]{\@secondoftwo}%
	\providecommand \href [0]{\begingroup \@sanitize@url \@href}%
	\providecommand \@href[1]{\@@startlink{#1}\@@href}%
	\providecommand \@@href[1]{\endgroup#1\@@endlink}%
	\providecommand \@sanitize@url [0]{\catcode `\\12\catcode `\$12\catcode
		`\&12\catcode `\#12\catcode `\^12\catcode `\_12\catcode `\%12\relax}%
	\providecommand \@@startlink[1]{}%
	\providecommand \@@endlink[0]{}%
	\providecommand \url  [0]{\begingroup\@sanitize@url \@url }%
	\providecommand \@url [1]{\endgroup\@href {#1}{\urlprefix }}%
	\providecommand \urlprefix  [0]{URL }%
	\providecommand \Eprint [0]{\href }%
	\providecommand \doibase [0]{https://doi.org/}%
	\providecommand \selectlanguage [0]{\@gobble}%
	\providecommand \bibinfo  [0]{\@secondoftwo}%
	\providecommand \bibfield  [0]{\@secondoftwo}%
	\providecommand \translation [1]{[#1]}%
	\providecommand \BibitemOpen [0]{}%
	\providecommand \bibitemStop [0]{}%
	\providecommand \bibitemNoStop [0]{.\EOS\space}%
	\providecommand \EOS [0]{\spacefactor3000\relax}%
	\providecommand \BibitemShut  [1]{\csname bibitem#1\endcsname}%
	\let\auto@bib@innerbib\@empty
	%</preamble>
	\bibitem [{\citenamefont {Tuval}\ \emph {et~al.}(2005)\citenamefont {Tuval},
		\citenamefont {Cisneros}, \citenamefont {Dombrowski}, \citenamefont
		{Wolgemuth}, \citenamefont {Kessler},\ and\ \citenamefont
		{Goldstein}}]{tuva:05}%
	\BibitemOpen
	\bibfield  {author} {\bibinfo {author} {\bibfnamefont {I.}~\bibnamefont
			{Tuval}}, \bibinfo {author} {\bibfnamefont {L.}~\bibnamefont {Cisneros}},
		\bibinfo {author} {\bibfnamefont {C.}~\bibnamefont {Dombrowski}}, \bibinfo
		{author} {\bibfnamefont {C.~W.}\ \bibnamefont {Wolgemuth}}, \bibinfo {author}
		{\bibfnamefont {J.~O.}\ \bibnamefont {Kessler}},\ and\ \bibinfo {author}
		{\bibfnamefont {R.~E.}\ \bibnamefont {Goldstein}},\ }\href
	{https://doi.org/10.1073/pnas.0406724102} {\bibfield  {journal} {\bibinfo
			{journal} {Proc. Natl. Acad. Sci.}\ }\textbf {\bibinfo {volume} {102}},\
		\bibinfo {pages} {2277} (\bibinfo {year} {2005})}\BibitemShut {NoStop}%
	\bibitem [{\citenamefont {Guasto}\ \emph {et~al.}(2010)\citenamefont {Guasto},
		\citenamefont {Johnson},\ and\ \citenamefont {Gollub}}]{guas:10}%
	\BibitemOpen
	\bibfield  {author} {\bibinfo {author} {\bibfnamefont {J.~S.}\ \bibnamefont
			{Guasto}}, \bibinfo {author} {\bibfnamefont {K.~A.}\ \bibnamefont
			{Johnson}},\ and\ \bibinfo {author} {\bibfnamefont {J.~P.}\ \bibnamefont
			{Gollub}},\ }\href {http://link.aps.org/doi/10.1103/PhysRevLett.105.168102}
	{\bibfield  {journal} {\bibinfo  {journal} {Phys. Rev. Lett.}\ }\textbf
		{\bibinfo {volume} {105}},\ \bibinfo {pages} {168102} (\bibinfo {year}
		{2010})}\BibitemShut {NoStop}%
	\bibitem [{\citenamefont {Ki{\o }rboe}\ \emph {et~al.}(2014)\citenamefont
		{Ki{\o }rboe}, \citenamefont {Jiang}, \citenamefont {Gon{\c{c}}alves},
		\citenamefont {Nielsen},\ and\ \citenamefont {Wadhwa}}]{kior:14}%
	\BibitemOpen
	\bibfield  {author} {\bibinfo {author} {\bibfnamefont {T.}~\bibnamefont
			{Ki{\o }rboe}}, \bibinfo {author} {\bibfnamefont {H.}~\bibnamefont {Jiang}},
		\bibinfo {author} {\bibfnamefont {R.~J.}\ \bibnamefont {Gon{\c{c}}alves}},
		\bibinfo {author} {\bibfnamefont {L.~T.}\ \bibnamefont {Nielsen}},\ and\
		\bibinfo {author} {\bibfnamefont {N.}~\bibnamefont {Wadhwa}},\ }\href
	{https://doi.org/10.1073/pnas.1405260111} {\bibfield  {journal} {\bibinfo
			{journal} {Proc. Natl. Acad. Sci. USA}\ }\textbf {\bibinfo {volume} {111}},\
		\bibinfo {pages} {11738} (\bibinfo {year} {2014})}\BibitemShut {NoStop}%
	\bibitem [{\citenamefont {Elgeti}\ \emph {et~al.}(2015)\citenamefont {Elgeti},
		\citenamefont {Winkler},\ and\ \citenamefont {Gompper}}]{elge:15}%
	\BibitemOpen
	\bibfield  {author} {\bibinfo {author} {\bibfnamefont {J.}~\bibnamefont
			{Elgeti}}, \bibinfo {author} {\bibfnamefont {R.~G.}\ \bibnamefont
			{Winkler}},\ and\ \bibinfo {author} {\bibfnamefont {G.}~\bibnamefont
			{Gompper}},\ }\href {https://doi.org/10.1088/0034-4885/78/5/056601}
	{\bibfield  {journal} {\bibinfo  {journal} {Rep. Prog. Phys.}\ }\textbf
		{\bibinfo {volume} {78}},\ \bibinfo {pages} {056601} (\bibinfo {year}
		{2015})}\BibitemShut {NoStop}%
	\bibitem [{\citenamefont {Bechinger}\ \emph {et~al.}(2016)\citenamefont
		{Bechinger}, \citenamefont {Di~Leonardo}, \citenamefont {L{\"o}wen},
		\citenamefont {Reichhardt}, \citenamefont {Volpe},\ and\ \citenamefont
		{Volpe}}]{bech:16}%
	\BibitemOpen
	\bibfield  {author} {\bibinfo {author} {\bibfnamefont {C.}~\bibnamefont
			{Bechinger}}, \bibinfo {author} {\bibfnamefont {R.}~\bibnamefont
			{Di~Leonardo}}, \bibinfo {author} {\bibfnamefont {H.}~\bibnamefont
			{L{\"o}wen}}, \bibinfo {author} {\bibfnamefont {C.}~\bibnamefont
			{Reichhardt}}, \bibinfo {author} {\bibfnamefont {G.}~\bibnamefont {Volpe}},\
		and\ \bibinfo {author} {\bibfnamefont {G.}~\bibnamefont {Volpe}},\ }\href
	{https://doi.org/10.1103/RevModPhys.88.045006} {\bibfield  {journal}
		{\bibinfo  {journal} {Rev. Mod. Phys.}\ }\textbf {\bibinfo {volume} {88}},\
		\bibinfo {pages} {045006} (\bibinfo {year} {2016})}\BibitemShut {NoStop}%
	\bibitem [{\citenamefont {China}\ and\ \citenamefont
		{Holzman}(2014)}]{chin:14}%
	\BibitemOpen
	\bibfield  {author} {\bibinfo {author} {\bibfnamefont {V.}~\bibnamefont
			{China}}\ and\ \bibinfo {author} {\bibfnamefont {R.}~\bibnamefont
			{Holzman}},\ }\href {https://doi.org/10.1073/pnas.1323205111} {\bibfield
		{journal} {\bibinfo  {journal} {Proc. Natl. Acad. Sci. USA}\ }\textbf
		{\bibinfo {volume} {111}},\ \bibinfo {pages} {8083} (\bibinfo {year}
		{2014})}\BibitemShut {NoStop}%
	\bibitem [{\citenamefont {Ki{\o}rboe}\ \emph {et~al.}(2009)\citenamefont
		{Ki{\o}rboe}, \citenamefont {Andersen}, \citenamefont {Langlois},
		\citenamefont {Jakobsen},\ and\ \citenamefont {Bohr}}]{kior:09}%
	\BibitemOpen
	\bibfield  {author} {\bibinfo {author} {\bibfnamefont {T.}~\bibnamefont
			{Ki{\o}rboe}}, \bibinfo {author} {\bibfnamefont {A.}~\bibnamefont
			{Andersen}}, \bibinfo {author} {\bibfnamefont {V.~J.}\ \bibnamefont
			{Langlois}}, \bibinfo {author} {\bibfnamefont {H.~H.}\ \bibnamefont
			{Jakobsen}},\ and\ \bibinfo {author} {\bibfnamefont {T.}~\bibnamefont
			{Bohr}},\ }\href {https://doi.org/10.1073/pnas.0903350106} {\bibfield
		{journal} {\bibinfo  {journal} {Proc. Natl. Acad. Sci. USA}\ }\textbf
		{\bibinfo {volume} {106}},\ \bibinfo {pages} {12394} (\bibinfo {year}
		{2009})}\BibitemShut {NoStop}%
	\bibitem [{\citenamefont {Theers}\ \emph {et~al.}(2018)\citenamefont {Theers},
		\citenamefont {Westphal}, \citenamefont {Qi}, \citenamefont {Winkler},\ and\
		\citenamefont {Gompper}}]{thee:18}%
	\BibitemOpen
	\bibfield  {author} {\bibinfo {author} {\bibfnamefont {M.}~\bibnamefont
			{Theers}}, \bibinfo {author} {\bibfnamefont {E.}~\bibnamefont {Westphal}},
		\bibinfo {author} {\bibfnamefont {K.}~\bibnamefont {Qi}}, \bibinfo {author}
		{\bibfnamefont {R.~G.}\ \bibnamefont {Winkler}},\ and\ \bibinfo {author}
		{\bibfnamefont {G.}~\bibnamefont {Gompper}},\ }\href
	{https://doi.org/10.1039/C8SM01390J} {\bibfield  {journal} {\bibinfo
			{journal} {Soft Matter}\ }\textbf {\bibinfo {volume} {14}},\ \bibinfo {pages}
		{8590} (\bibinfo {year} {2018})}\BibitemShut {NoStop}%
	\bibitem [{\citenamefont {Qi}\ \emph {et~al.}(2022)\citenamefont {Qi},
		\citenamefont {Westphal}, \citenamefont {Gompper},\ and\ \citenamefont
		{Winkler}}]{qi:22}%
	\BibitemOpen
	\bibfield  {author} {\bibinfo {author} {\bibfnamefont {K.}~\bibnamefont
			{Qi}}, \bibinfo {author} {\bibfnamefont {E.}~\bibnamefont {Westphal}},
		\bibinfo {author} {\bibfnamefont {G.}~\bibnamefont {Gompper}},\ and\ \bibinfo
		{author} {\bibfnamefont {R.~G.}\ \bibnamefont {Winkler}},\ }\href
	{https://doi.org/10.1038/s42005-022-00820-7} {\bibfield  {journal} {\bibinfo
			{journal} {Commun. Phys.}\ }\textbf {\bibinfo {volume} {5}},\ \bibinfo
		{pages} {49} (\bibinfo {year} {2022})}\BibitemShut {NoStop}%
	\bibitem [{\citenamefont {Samatas}\ and\ \citenamefont
		{Lintuvuori}(2023)}]{sama:23}%
	\BibitemOpen
	\bibfield  {author} {\bibinfo {author} {\bibfnamefont {S.}~\bibnamefont
			{Samatas}}\ and\ \bibinfo {author} {\bibfnamefont {J.}~\bibnamefont
			{Lintuvuori}},\ }\href {https://doi.org/10.1103/PhysRevLett.130.024001}
	{\bibfield  {journal} {\bibinfo  {journal} {Phys. Rev. Lett.}\ }\textbf
		{\bibinfo {volume} {130}},\ \bibinfo {pages} {024001} (\bibinfo {year}
		{2023})}\BibitemShut {NoStop}%
	\bibitem [{\citenamefont {Wensink}\ \emph {et~al.}(2012)\citenamefont
		{Wensink}, \citenamefont {Dunkel}, \citenamefont {Heidenreich}, \citenamefont
		{Drescher}, \citenamefont {Goldstein}, \citenamefont {L{\"o}wen},\ and\
		\citenamefont {Yeomans}}]{wens:12}%
	\BibitemOpen
	\bibfield  {author} {\bibinfo {author} {\bibfnamefont {H.~H.}\ \bibnamefont
			{Wensink}}, \bibinfo {author} {\bibfnamefont {J.}~\bibnamefont {Dunkel}},
		\bibinfo {author} {\bibfnamefont {S.}~\bibnamefont {Heidenreich}}, \bibinfo
		{author} {\bibfnamefont {K.}~\bibnamefont {Drescher}}, \bibinfo {author}
		{\bibfnamefont {R.~E.}\ \bibnamefont {Goldstein}}, \bibinfo {author}
		{\bibfnamefont {H.}~\bibnamefont {L{\"o}wen}},\ and\ \bibinfo {author}
		{\bibfnamefont {J.~M.}\ \bibnamefont {Yeomans}},\ }\href
	{https://doi.org/10.1073/pnas.1202032109} {\bibfield  {journal} {\bibinfo
			{journal} {Proc. Natl. Acad. Sci. USA}\ }\textbf {\bibinfo {volume} {109}},\
		\bibinfo {pages} {14308} (\bibinfo {year} {2012})}\BibitemShut {NoStop}%
	\bibitem [{\citenamefont {Dunkel}\ \emph {et~al.}(2013)\citenamefont {Dunkel},
		\citenamefont {Heidenreich}, \citenamefont {Drescher}, \citenamefont
		{Wensink}, \citenamefont {B{\"a}r},\ and\ \citenamefont
		{Goldstein}}]{dunk:13}%
	\BibitemOpen
	\bibfield  {author} {\bibinfo {author} {\bibfnamefont {J.}~\bibnamefont
			{Dunkel}}, \bibinfo {author} {\bibfnamefont {S.}~\bibnamefont {Heidenreich}},
		\bibinfo {author} {\bibfnamefont {K.}~\bibnamefont {Drescher}}, \bibinfo
		{author} {\bibfnamefont {H.~H.}\ \bibnamefont {Wensink}}, \bibinfo {author}
		{\bibfnamefont {M.}~\bibnamefont {B{\"a}r}},\ and\ \bibinfo {author}
		{\bibfnamefont {R.~E.}\ \bibnamefont {Goldstein}},\ }\href
	{https://doi.org/10.1103/PhysRevLett.110.228102} {\bibfield  {journal}
		{\bibinfo  {journal} {Phys. Rev. Lett.}\ }\textbf {\bibinfo {volume} {110}},\
		\bibinfo {pages} {228102} (\bibinfo {year} {2013})}\BibitemShut {NoStop}%
	\bibitem [{\citenamefont {Aranson}(2022)}]{aran:22}%
	\BibitemOpen
	\bibfield  {author} {\bibinfo {author} {\bibfnamefont {I.~S.}\ \bibnamefont
			{Aranson}},\ }\href {https://doi.org/10.1088/1361-6633/ac723d} {\bibfield
		{journal} {\bibinfo  {journal} {Rep. Prog. Phys.}\ }\textbf {\bibinfo
			{volume} {85}},\ \bibinfo {pages} {076601} (\bibinfo {year}
		{2022})}\BibitemShut {NoStop}%
	\bibitem [{\citenamefont {Ricotti}\ \emph {et~al.}(2017)\citenamefont
		{Ricotti}, \citenamefont {Trimmer}, \citenamefont {Feinberg}, \citenamefont
		{Raman}, \citenamefont {Parker}, \citenamefont {Bashir}, \citenamefont
		{Sitti}, \citenamefont {Martel}, \citenamefont {Dario},\ and\ \citenamefont
		{Menciassi}}]{rico:17}%
	\BibitemOpen
	\bibfield  {author} {\bibinfo {author} {\bibfnamefont {L.}~\bibnamefont
			{Ricotti}}, \bibinfo {author} {\bibfnamefont {B.}~\bibnamefont {Trimmer}},
		\bibinfo {author} {\bibfnamefont {A.~W.}\ \bibnamefont {Feinberg}}, \bibinfo
		{author} {\bibfnamefont {R.}~\bibnamefont {Raman}}, \bibinfo {author}
		{\bibfnamefont {K.~K.}\ \bibnamefont {Parker}}, \bibinfo {author}
		{\bibfnamefont {R.}~\bibnamefont {Bashir}}, \bibinfo {author} {\bibfnamefont
			{M.}~\bibnamefont {Sitti}}, \bibinfo {author} {\bibfnamefont
			{S.}~\bibnamefont {Martel}}, \bibinfo {author} {\bibfnamefont
			{P.}~\bibnamefont {Dario}},\ and\ \bibinfo {author} {\bibfnamefont
			{A.}~\bibnamefont {Menciassi}},\ }\href
	{https://doi.org/10.1126/scirobotics.aaq0495} {\bibfield  {journal} {\bibinfo
			{journal} {Sci. Robotics}\ }\textbf {\bibinfo {volume} {2}},\ \bibinfo
		{pages} {eaaq0495} (\bibinfo {year} {2017})}\BibitemShut {NoStop}%
	\bibitem [{\citenamefont {Huang}\ \emph {et~al.}(2022)\citenamefont {Huang},
		\citenamefont {Gu},\ and\ \citenamefont {Nelson}}]{huan:22}%
	\BibitemOpen
	\bibfield  {author} {\bibinfo {author} {\bibfnamefont {T.-Y.}\ \bibnamefont
			{Huang}}, \bibinfo {author} {\bibfnamefont {H.}~\bibnamefont {Gu}},\ and\
		\bibinfo {author} {\bibfnamefont {B.~J.}\ \bibnamefont {Nelson}},\ }\href
	{https://doi.org/10.1146/annurev-control-042920-013322} {\bibfield  {journal}
		{\bibinfo  {journal} {Annu. Rev. Control Robot. Auton. Syst.}\ }\textbf
		{\bibinfo {volume} {5}},\ \bibinfo {pages} {279} (\bibinfo {year}
		{2022})}\BibitemShut {NoStop}%
	\bibitem [{\citenamefont {B{\"a}uerle}\ \emph {et~al.}(2018)\citenamefont
		{B{\"a}uerle}, \citenamefont {Fischer}, \citenamefont {Speck},\ and\
		\citenamefont {Bechinger}}]{baeu:18}%
	\BibitemOpen
	\bibfield  {author} {\bibinfo {author} {\bibfnamefont {T.}~\bibnamefont
			{B{\"a}uerle}}, \bibinfo {author} {\bibfnamefont {A.}~\bibnamefont
			{Fischer}}, \bibinfo {author} {\bibfnamefont {T.}~\bibnamefont {Speck}},\
		and\ \bibinfo {author} {\bibfnamefont {C.}~\bibnamefont {Bechinger}},\ }\href
	{https://doi.org/10.1038/s41467-018-05675-7} {\bibfield  {journal} {\bibinfo
			{journal} {Nat. Commun.}\ }\textbf {\bibinfo {volume} {9}},\ \bibinfo {pages}
		{3232} (\bibinfo {year} {2018})}\BibitemShut {NoStop}%
	\bibitem [{\citenamefont {Selmke}\ \emph {et~al.}(2018)\citenamefont {Selmke},
		\citenamefont {Khadka}, \citenamefont {Bregulla}, \citenamefont {Cichos},\
		and\ \citenamefont {Yang}}]{selm:18}%
	\BibitemOpen
	\bibfield  {author} {\bibinfo {author} {\bibfnamefont {M.}~\bibnamefont
			{Selmke}}, \bibinfo {author} {\bibfnamefont {U.}~\bibnamefont {Khadka}},
		\bibinfo {author} {\bibfnamefont {A.~P.}\ \bibnamefont {Bregulla}}, \bibinfo
		{author} {\bibfnamefont {F.}~\bibnamefont {Cichos}},\ and\ \bibinfo {author}
		{\bibfnamefont {H.}~\bibnamefont {Yang}},\ }\href
	{https://doi.org/10.1039/C7CP06559K} {\bibfield  {journal} {\bibinfo
			{journal} {Phys. Chem. Chem. Phys.}\ }\textbf {\bibinfo {volume} {20}},\
		\bibinfo {pages} {10502} (\bibinfo {year} {2018})}\BibitemShut {NoStop}%
	\bibitem [{\citenamefont {Lavergne}\ \emph {et~al.}(2019)\citenamefont
		{Lavergne}, \citenamefont {Wendehenne}, \citenamefont {B{\"a}uerle},\ and\
		\citenamefont {Bechinger}}]{lave:19}%
	\BibitemOpen
	\bibfield  {author} {\bibinfo {author} {\bibfnamefont {F.~A.}\ \bibnamefont
			{Lavergne}}, \bibinfo {author} {\bibfnamefont {H.}~\bibnamefont
			{Wendehenne}}, \bibinfo {author} {\bibfnamefont {T.}~\bibnamefont
			{B{\"a}uerle}},\ and\ \bibinfo {author} {\bibfnamefont {C.}~\bibnamefont
			{Bechinger}},\ }\href {https://doi.org/10.1126/science.aau5347} {\bibfield
		{journal} {\bibinfo  {journal} {Science}\ }\textbf {\bibinfo {volume}
			{364}},\ \bibinfo {pages} {70} (\bibinfo {year} {2019})}\BibitemShut
	{NoStop}%
	\bibitem [{\citenamefont {Kaspar}\ \emph {et~al.}(2021)\citenamefont {Kaspar},
		\citenamefont {Ravoo}, \citenamefont {van~der Wiel}, \citenamefont {Wegner},\
		and\ \citenamefont {Pernice}}]{kasp:21}%
	\BibitemOpen
	\bibfield  {author} {\bibinfo {author} {\bibfnamefont {C.}~\bibnamefont
			{Kaspar}}, \bibinfo {author} {\bibfnamefont {B.~J.}\ \bibnamefont {Ravoo}},
		\bibinfo {author} {\bibfnamefont {W.~G.}\ \bibnamefont {van~der Wiel}},
		\bibinfo {author} {\bibfnamefont {S.~V.}\ \bibnamefont {Wegner}},\ and\
		\bibinfo {author} {\bibfnamefont {W.~H.~P.}\ \bibnamefont {Pernice}},\ }\href
	{https://doi.org/10.1038/s41586-021-03453-y} {\bibfield  {journal} {\bibinfo
			{journal} {Nature}\ }\textbf {\bibinfo {volume} {594}},\ \bibinfo {pages}
		{345} (\bibinfo {year} {2021})}\BibitemShut {NoStop}%
	\bibitem [{\citenamefont {Erkoc}\ \emph {et~al.}(2019)\citenamefont {Erkoc},
		\citenamefont {Yasa}, \citenamefont {Ceylan}, \citenamefont {Yasa},
		\citenamefont {Alapan},\ and\ \citenamefont {Sitti}}]{erko:19}%
	\BibitemOpen
	\bibfield  {author} {\bibinfo {author} {\bibfnamefont {P.}~\bibnamefont
			{Erkoc}}, \bibinfo {author} {\bibfnamefont {I.~C.}\ \bibnamefont {Yasa}},
		\bibinfo {author} {\bibfnamefont {H.}~\bibnamefont {Ceylan}}, \bibinfo
		{author} {\bibfnamefont {O.}~\bibnamefont {Yasa}}, \bibinfo {author}
		{\bibfnamefont {Y.}~\bibnamefont {Alapan}},\ and\ \bibinfo {author}
		{\bibfnamefont {M.}~\bibnamefont {Sitti}},\ }\href
	{https://doi.org/https://doi.org/10.1002/adtp.201800064} {\bibfield
		{journal} {\bibinfo  {journal} {Adv. Therap.}\ }\textbf {\bibinfo {volume}
			{2}},\ \bibinfo {pages} {1800064} (\bibinfo {year} {2019})}\BibitemShut
	{NoStop}%
	\bibitem [{\citenamefont {Kurzthaler}\ \emph {et~al.}(2021)\citenamefont
		{Kurzthaler}, \citenamefont {Mandal}, \citenamefont {Bhattacharjee},
		\citenamefont {L{\"o}wen}, \citenamefont {Datta},\ and\ \citenamefont
		{Stone}}]{kurz:21}%
	\BibitemOpen
	\bibfield  {author} {\bibinfo {author} {\bibfnamefont {C.}~\bibnamefont
			{Kurzthaler}}, \bibinfo {author} {\bibfnamefont {S.}~\bibnamefont {Mandal}},
		\bibinfo {author} {\bibfnamefont {T.}~\bibnamefont {Bhattacharjee}}, \bibinfo
		{author} {\bibfnamefont {H.}~\bibnamefont {L{\"o}wen}}, \bibinfo {author}
		{\bibfnamefont {S.~S.}\ \bibnamefont {Datta}},\ and\ \bibinfo {author}
		{\bibfnamefont {H.~A.}\ \bibnamefont {Stone}},\ }\href
	{https://doi.org/10.1038/s41467-021-26942-0} {\bibfield  {journal} {\bibinfo
			{journal} {Nat. Commun.}\ }\textbf {\bibinfo {volume} {12}},\ \bibinfo
		{pages} {7088} (\bibinfo {year} {2021})}\BibitemShut {NoStop}%
	\bibitem [{\citenamefont {Cvetkovic}\ \emph {et~al.}(2014)\citenamefont
		{Cvetkovic}, \citenamefont {Raman}, \citenamefont {Chan}, \citenamefont
		{Williams}, \citenamefont {Tolish}, \citenamefont {Bajaj}, \citenamefont
		{Sakar}, \citenamefont {Asada}, \citenamefont {Saif},\ and\ \citenamefont
		{Bashir}}]{cvet:14}%
	\BibitemOpen
	\bibfield  {author} {\bibinfo {author} {\bibfnamefont {C.}~\bibnamefont
			{Cvetkovic}}, \bibinfo {author} {\bibfnamefont {R.}~\bibnamefont {Raman}},
		\bibinfo {author} {\bibfnamefont {V.}~\bibnamefont {Chan}}, \bibinfo {author}
		{\bibfnamefont {B.~J.}\ \bibnamefont {Williams}}, \bibinfo {author}
		{\bibfnamefont {M.}~\bibnamefont {Tolish}}, \bibinfo {author} {\bibfnamefont
			{P.}~\bibnamefont {Bajaj}}, \bibinfo {author} {\bibfnamefont {M.~S.}\
			\bibnamefont {Sakar}}, \bibinfo {author} {\bibfnamefont {H.~H.}\ \bibnamefont
			{Asada}}, \bibinfo {author} {\bibfnamefont {M.~T.~A.}\ \bibnamefont {Saif}},\
		and\ \bibinfo {author} {\bibfnamefont {R.}~\bibnamefont {Bashir}},\ }\href
	{https://doi.org/10.1073/pnas.1401577111} {\bibfield  {journal} {\bibinfo
			{journal} {Proc. Natl. Acad. Sci. USA}\ }\textbf {\bibinfo {volume} {111}},\
		\bibinfo {pages} {10125} (\bibinfo {year} {2014})}\BibitemShut {NoStop}%
	\bibitem [{\citenamefont {Dai}\ \emph {et~al.}(2016)\citenamefont {Dai},
		\citenamefont {Wang}, \citenamefont {Xiong}, \citenamefont {Zhan},
		\citenamefont {Dai}, \citenamefont {Li}, \citenamefont {Feng},\ and\
		\citenamefont {Tang}}]{dai:16}%
	\BibitemOpen
	\bibfield  {author} {\bibinfo {author} {\bibfnamefont {B.}~\bibnamefont
			{Dai}}, \bibinfo {author} {\bibfnamefont {J.}~\bibnamefont {Wang}}, \bibinfo
		{author} {\bibfnamefont {Z.}~\bibnamefont {Xiong}}, \bibinfo {author}
		{\bibfnamefont {X.}~\bibnamefont {Zhan}}, \bibinfo {author} {\bibfnamefont
			{W.}~\bibnamefont {Dai}}, \bibinfo {author} {\bibfnamefont {C.-C.}\
			\bibnamefont {Li}}, \bibinfo {author} {\bibfnamefont {S.-P.}\ \bibnamefont
			{Feng}},\ and\ \bibinfo {author} {\bibfnamefont {J.}~\bibnamefont {Tang}},\
	}\href {https://doi.org/10.1038/nnano.2016.187} {\bibfield  {journal}
		{\bibinfo  {journal} {Nat. Nanotechnol.}\ }\textbf {\bibinfo {volume} {11}},\
		\bibinfo {pages} {1087} (\bibinfo {year} {2016})}\BibitemShut {NoStop}%
	\bibitem [{\citenamefont {Tsang}\ \emph {et~al.}(2018)\citenamefont {Tsang},
		\citenamefont {Lam},\ and\ \citenamefont {Riedel-Kruse}}]{Tsan:18}%
	\BibitemOpen
	\bibfield  {author} {\bibinfo {author} {\bibfnamefont {A.~C.~H.}\
			\bibnamefont {Tsang}}, \bibinfo {author} {\bibfnamefont {A.~T.}\ \bibnamefont
			{Lam}},\ and\ \bibinfo {author} {\bibfnamefont {I.~H.}\ \bibnamefont
			{Riedel-Kruse}},\ }\href {https://doi.org/10.1038/s41567-018-0277-7}
	{\bibfield  {journal} {\bibinfo  {journal} {Nat. Phys.}\ }\textbf {\bibinfo
			{volume} {14}},\ \bibinfo {pages} {1216} (\bibinfo {year}
		{2018})}\BibitemShut {NoStop}%
	\bibitem [{\citenamefont {Alvarez}\ \emph {et~al.}(2021)\citenamefont
		{Alvarez}, \citenamefont {Fernandez-Rodriguez}, \citenamefont {Alegria},
		\citenamefont {Arrese-Igor}, \citenamefont {Zhao}, \citenamefont
		{Kr{\"o}ger},\ and\ \citenamefont {Isa}}]{alva:21}%
	\BibitemOpen
	\bibfield  {author} {\bibinfo {author} {\bibfnamefont {L.}~\bibnamefont
			{Alvarez}}, \bibinfo {author} {\bibfnamefont {M.~A.}\ \bibnamefont
			{Fernandez-Rodriguez}}, \bibinfo {author} {\bibfnamefont {A.}~\bibnamefont
			{Alegria}}, \bibinfo {author} {\bibfnamefont {S.}~\bibnamefont
			{Arrese-Igor}}, \bibinfo {author} {\bibfnamefont {K.}~\bibnamefont {Zhao}},
		\bibinfo {author} {\bibfnamefont {M.}~\bibnamefont {Kr{\"o}ger}},\ and\
		\bibinfo {author} {\bibfnamefont {L.}~\bibnamefont {Isa}},\ }\href
	{https://doi.org/10.1038/s41467-021-25108-2} {\bibfield  {journal} {\bibinfo
			{journal} {Nat. Commun.}\ }\textbf {\bibinfo {volume} {12}},\ \bibinfo
		{pages} {4762} (\bibinfo {year} {2021})}\BibitemShut {NoStop}%
	\bibitem [{\citenamefont {Lighthill}(1952)}]{ligh:52}%
	\BibitemOpen
	\bibfield  {author} {\bibinfo {author} {\bibfnamefont {M.~J.}\ \bibnamefont
			{Lighthill}},\ }\href {https://doi.org/10.1002/cpa.3160050201} {\bibfield
		{journal} {\bibinfo  {journal} {Comm. Pure Appl. Math.}\ }\textbf {\bibinfo
			{volume} {5}},\ \bibinfo {pages} {109} (\bibinfo {year} {1952})}\BibitemShut
	{NoStop}%
	\bibitem [{\citenamefont {Blake}(1971)}]{blak:71}%
	\BibitemOpen
	\bibfield  {author} {\bibinfo {author} {\bibfnamefont {J.~R.}\ \bibnamefont
			{Blake}},\ }\href {https://doi.org/10.1017/S002211207100048X} {\bibfield
		{journal} {\bibinfo  {journal} {J. Fluid Mech.}\ }\textbf {\bibinfo {volume}
			{46}},\ \bibinfo {pages} {199} (\bibinfo {year} {1971})}\BibitemShut
	{NoStop}%
	\bibitem [{\citenamefont {Pak}\ and\ \citenamefont {Lauga}(2014)}]{pak:14}%
	\BibitemOpen
	\bibfield  {author} {\bibinfo {author} {\bibfnamefont {O.~S.}\ \bibnamefont
			{Pak}}\ and\ \bibinfo {author} {\bibfnamefont {E.}~\bibnamefont {Lauga}},\
	}\href {https://doi.org/10.1007/s10665-014-9690-9} {\bibfield  {journal}
		{\bibinfo  {journal} {J. Eng. Math.}\ }\textbf {\bibinfo {volume} {88}},\
		\bibinfo {pages} {1} (\bibinfo {year} {2014})}\BibitemShut {NoStop}%
	\bibitem [{\citenamefont {Barberis}\ and\ \citenamefont
		{Peruani}(2016)}]{barb:16}%
	\BibitemOpen
	\bibfield  {author} {\bibinfo {author} {\bibfnamefont {L.}~\bibnamefont
			{Barberis}}\ and\ \bibinfo {author} {\bibfnamefont {F.}~\bibnamefont
			{Peruani}},\ }\href {https://doi.org/10.1103/PhysRevLett.117.248001}
	{\bibfield  {journal} {\bibinfo  {journal} {Phys. Rev. Lett.}\ }\textbf
		{\bibinfo {volume} {117}},\ \bibinfo {pages} {248001} (\bibinfo {year}
		{2016})}\BibitemShut {NoStop}%
	\bibitem [{\citenamefont {Goh}\ \emph {et~al.}(2022)\citenamefont {Goh},
		\citenamefont {Winkler},\ and\ \citenamefont {Gompper}}]{goh:22}%
	\BibitemOpen
	\bibfield  {author} {\bibinfo {author} {\bibfnamefont {S.}~\bibnamefont
			{Goh}}, \bibinfo {author} {\bibfnamefont {R.~G.}\ \bibnamefont {Winkler}},\
		and\ \bibinfo {author} {\bibfnamefont {G.}~\bibnamefont {Gompper}},\ }\href
	{https://doi.org/10.1088/1367-2630/ac924f} {\bibfield  {journal} {\bibinfo
			{journal} {New J. Phys.}\ }\textbf {\bibinfo {volume} {24}},\ \bibinfo
		{pages} {093039} (\bibinfo {year} {2022})}\BibitemShut {NoStop}%
	\bibitem [{\citenamefont {Vicsek}\ \emph {et~al.}(1995)\citenamefont {Vicsek},
		\citenamefont {Czir{\'o}k}, \citenamefont {Ben-Jacob}, \citenamefont
		{Cohen},\ and\ \citenamefont {Shochet}}]{vics:95}%
	\BibitemOpen
	\bibfield  {author} {\bibinfo {author} {\bibfnamefont {T.}~\bibnamefont
			{Vicsek}}, \bibinfo {author} {\bibfnamefont {A.}~\bibnamefont {Czir{\'o}k}},
		\bibinfo {author} {\bibfnamefont {E.}~\bibnamefont {Ben-Jacob}}, \bibinfo
		{author} {\bibfnamefont {I.}~\bibnamefont {Cohen}},\ and\ \bibinfo {author}
		{\bibfnamefont {O.}~\bibnamefont {Shochet}},\ }\href
	{https://doi.org/10.1103/PhysRevLett.75.1226} {\bibfield  {journal} {\bibinfo
			{journal} {Phys. Rev. Lett.}\ }\textbf {\bibinfo {volume} {75}},\ \bibinfo
		{pages} {1226} (\bibinfo {year} {1995})}\BibitemShut {NoStop}%
	\bibitem [{\citenamefont {Chat{\'e}}(2020)}]{chat:20}%
	\BibitemOpen
	\bibfield  {author} {\bibinfo {author} {\bibfnamefont {H.}~\bibnamefont
			{Chat{\'e}}},\ }\href
	{https://doi.org/10.1146/annurev-conmatphys-031119-050752} {\bibfield
		{journal} {\bibinfo  {journal} {Annu. Rev. Condens. Matter Phys.}\ }\textbf
		{\bibinfo {volume} {11}},\ \bibinfo {pages} {189} (\bibinfo {year}
		{2020})}\BibitemShut {NoStop}%
	\bibitem [{\citenamefont {Chepizhko}\ \emph {et~al.}(2021)\citenamefont
		{Chepizhko}, \citenamefont {Saintillan},\ and\ \citenamefont
		{Peruani}}]{chep:21}%
	\BibitemOpen
	\bibfield  {author} {\bibinfo {author} {\bibfnamefont {O.}~\bibnamefont
			{Chepizhko}}, \bibinfo {author} {\bibfnamefont {D.}~\bibnamefont
			{Saintillan}},\ and\ \bibinfo {author} {\bibfnamefont {F.}~\bibnamefont
			{Peruani}},\ }\href {https://doi.org/10.1039/D0SM01220C} {\bibfield
		{journal} {\bibinfo  {journal} {Soft Matter}\ }\textbf {\bibinfo {volume}
			{17}},\ \bibinfo {pages} {3113} (\bibinfo {year} {2021})}\BibitemShut
	{NoStop}%
	\bibitem [{\citenamefont {Ishikawa}\ \emph {et~al.}(2006)\citenamefont
		{Ishikawa}, \citenamefont {Simmonds},\ and\ \citenamefont
		{Pedley}}]{ishi:06}%
	\BibitemOpen
	\bibfield  {author} {\bibinfo {author} {\bibfnamefont {T.}~\bibnamefont
			{Ishikawa}}, \bibinfo {author} {\bibfnamefont {M.~P.}\ \bibnamefont
			{Simmonds}},\ and\ \bibinfo {author} {\bibfnamefont {T.~J.}\ \bibnamefont
			{Pedley}},\ }\href {https://doi.org/10.1017/S0022112006002631} {\bibfield
		{journal} {\bibinfo  {journal} {J. Fluid Mech.}\ }\textbf {\bibinfo {volume}
			{568}},\ \bibinfo {pages} {119} (\bibinfo {year} {2006})}\BibitemShut
	{NoStop}%
	\bibitem [{\citenamefont {Shaebani}\ \emph {et~al.}(2020)\citenamefont
		{Shaebani}, \citenamefont {Wysocki}, \citenamefont {Winkler}, \citenamefont
		{Gompper},\ and\ \citenamefont {Rieger}}]{shae:20}%
	\BibitemOpen
	\bibfield  {author} {\bibinfo {author} {\bibfnamefont {M.~R.}\ \bibnamefont
			{Shaebani}}, \bibinfo {author} {\bibfnamefont {A.}~\bibnamefont {Wysocki}},
		\bibinfo {author} {\bibfnamefont {R.~G.}\ \bibnamefont {Winkler}}, \bibinfo
		{author} {\bibfnamefont {G.}~\bibnamefont {Gompper}},\ and\ \bibinfo {author}
		{\bibfnamefont {H.}~\bibnamefont {Rieger}},\ }\href
	{https://doi.org/10.1038/s42254-020-0152-1} {\bibfield  {journal} {\bibinfo
			{journal} {Nat. Rev. Phys.}\ }\textbf {\bibinfo {volume} {2}},\ \bibinfo
		{pages} {181} (\bibinfo {year} {2020})}\BibitemShut {NoStop}%
	\bibitem [{\citenamefont {Kapral}(2008)}]{kapr:08}%
	\BibitemOpen
	\bibfield  {author} {\bibinfo {author} {\bibfnamefont {R.}~\bibnamefont
			{Kapral}},\ }\href {https://doi.org/10.1002/9780470371572.ch2} {\bibfield
		{journal} {\bibinfo  {journal} {Adv. Chem. Phys.}\ }\textbf {\bibinfo
			{volume} {140}},\ \bibinfo {pages} {89} (\bibinfo {year} {2008})}\BibitemShut
	{NoStop}%
	\bibitem [{\citenamefont {Gompper}\ \emph {et~al.}(2009)\citenamefont
		{Gompper}, \citenamefont {Ihle}, \citenamefont {Kroll},\ and\ \citenamefont
		{Winkler}}]{gomp:09}%
	\BibitemOpen
	\bibfield  {author} {\bibinfo {author} {\bibfnamefont {G.}~\bibnamefont
			{Gompper}}, \bibinfo {author} {\bibfnamefont {T.}~\bibnamefont {Ihle}},
		\bibinfo {author} {\bibfnamefont {D.~M.}\ \bibnamefont {Kroll}},\ and\
		\bibinfo {author} {\bibfnamefont {R.~G.}\ \bibnamefont {Winkler}},\ }\href
	{https://doi.org/10.1007/978-3-540-87706-6{\_1}} {\bibfield  {journal}
		{\bibinfo  {journal} {Adv. Polym. Sci.}\ }\textbf {\bibinfo {volume} {221}},\
		\bibinfo {pages} {1} (\bibinfo {year} {2009})}\BibitemShut {NoStop}%
	\bibitem [{\citenamefont {Huang}\ \emph {et~al.}(2012)\citenamefont {Huang},
		\citenamefont {Gompper},\ and\ \citenamefont {Winkler}}]{huan:12}%
	\BibitemOpen
	\bibfield  {author} {\bibinfo {author} {\bibfnamefont {C.-C.}\ \bibnamefont
			{Huang}}, \bibinfo {author} {\bibfnamefont {G.}~\bibnamefont {Gompper}},\
		and\ \bibinfo {author} {\bibfnamefont {R.~G.}\ \bibnamefont {Winkler}},\
	}\href {https://doi.org/10.1103/PhysRevE.86.056711} {\bibfield  {journal}
		{\bibinfo  {journal} {Phys. Rev. E}\ }\textbf {\bibinfo {volume} {86}},\
		\bibinfo {pages} {056711} (\bibinfo {year} {2012})}\BibitemShut {NoStop}%
	\bibitem [{\citenamefont {Noguchi}\ \emph {et~al.}(2007)\citenamefont
		{Noguchi}, \citenamefont {Kikuchi},\ and\ \citenamefont {Gompper}}]{nogu:07}%
	\BibitemOpen
	\bibfield  {author} {\bibinfo {author} {\bibfnamefont {H.}~\bibnamefont
			{Noguchi}}, \bibinfo {author} {\bibfnamefont {N.}~\bibnamefont {Kikuchi}},\
		and\ \bibinfo {author} {\bibfnamefont {G.}~\bibnamefont {Gompper}},\
	}\href@noop {} {\bibfield  {journal} {\bibinfo  {journal} {EPL}\ }\textbf
		{\bibinfo {volume} {78}},\ \bibinfo {pages} {10005} (\bibinfo {year}
		{2007})}\BibitemShut {NoStop}%
	\bibitem [{\citenamefont {Theers}\ \emph
		{et~al.}(2016{\natexlab{a}})\citenamefont {Theers}, \citenamefont {Westphal},
		\citenamefont {Gompper},\ and\ \citenamefont {Winkler}}]{thee:16}%
	\BibitemOpen
	\bibfield  {author} {\bibinfo {author} {\bibfnamefont {M.}~\bibnamefont
			{Theers}}, \bibinfo {author} {\bibfnamefont {E.}~\bibnamefont {Westphal}},
		\bibinfo {author} {\bibfnamefont {G.}~\bibnamefont {Gompper}},\ and\ \bibinfo
		{author} {\bibfnamefont {R.~G.}\ \bibnamefont {Winkler}},\ }\href
	{https://doi.org/10.1103/PhysRevE.93.032604} {\bibfield  {journal} {\bibinfo
			{journal} {Phys. Rev. E}\ }\textbf {\bibinfo {volume} {93}},\ \bibinfo
		{pages} {032604} (\bibinfo {year} {2016}{\natexlab{a}})}\BibitemShut
	{NoStop}%
	\bibitem [{\citenamefont {Hu}\ \emph {et~al.}(2015{\natexlab{a}})\citenamefont
		{Hu}, \citenamefont {Yang}, \citenamefont {Gompper},\ and\ \citenamefont
		{Winkler}}]{hu:15.1}%
	\BibitemOpen
	\bibfield  {author} {\bibinfo {author} {\bibfnamefont {J.}~\bibnamefont
			{Hu}}, \bibinfo {author} {\bibfnamefont {M.}~\bibnamefont {Yang}}, \bibinfo
		{author} {\bibfnamefont {G.}~\bibnamefont {Gompper}},\ and\ \bibinfo {author}
		{\bibfnamefont {R.~G.}\ \bibnamefont {Winkler}},\ }\href
	{https://doi.org/10.1039/C5SM01678A} {\bibfield  {journal} {\bibinfo
			{journal} {Soft Matter}\ }\textbf {\bibinfo {volume} {11}},\ \bibinfo {pages}
		{7867} (\bibinfo {year} {2015}{\natexlab{a}})}\BibitemShut {NoStop}%
	\bibitem [{\citenamefont {Llopis}\ and\ \citenamefont
		{Pagonabarraga}(2010)}]{llop:10}%
	\BibitemOpen
	\bibfield  {author} {\bibinfo {author} {\bibfnamefont {I.}~\bibnamefont
			{Llopis}}\ and\ \bibinfo {author} {\bibfnamefont {I.}~\bibnamefont
			{Pagonabarraga}},\ }\href {https://doi.org/10.1016/j.jnnfm.2010.01.023}
	{\bibfield  {journal} {\bibinfo  {journal} {J. Non-Newtonian Fluid Mech.}\
		}\textbf {\bibinfo {volume} {165}},\ \bibinfo {pages} {946} (\bibinfo {year}
		{2010})}\BibitemShut {NoStop}%
	\bibitem [{\citenamefont {G{\"o}tze}\ and\ \citenamefont
		{Gompper}(2010)}]{goet:10}%
	\BibitemOpen
	\bibfield  {author} {\bibinfo {author} {\bibfnamefont {I.~O.}\ \bibnamefont
			{G{\"o}tze}}\ and\ \bibinfo {author} {\bibfnamefont {G.}~\bibnamefont
			{Gompper}},\ }\href {https://doi.org/10.1103/PhysRevE.82.041921} {\bibfield
		{journal} {\bibinfo  {journal} {Phys. Rev. E}\ }\textbf {\bibinfo {volume}
			{82}},\ \bibinfo {pages} {041921} (\bibinfo {year} {2010})}\BibitemShut
	{NoStop}%
	\bibitem [{\citenamefont {Theers}\ \emph
		{et~al.}(2016{\natexlab{b}})\citenamefont {Theers}, \citenamefont {Westphal},
		\citenamefont {Gompper},\ and\ \citenamefont {Winkler}}]{thee:16.1}%
	\BibitemOpen
	\bibfield  {author} {\bibinfo {author} {\bibfnamefont {M.}~\bibnamefont
			{Theers}}, \bibinfo {author} {\bibfnamefont {E.}~\bibnamefont {Westphal}},
		\bibinfo {author} {\bibfnamefont {G.}~\bibnamefont {Gompper}},\ and\ \bibinfo
		{author} {\bibfnamefont {R.~G.}\ \bibnamefont {Winkler}},\ }\href
	{https://doi.org/10.1039/C6SM01424K} {\bibfield  {journal} {\bibinfo
			{journal} {Soft Matter}\ }\textbf {\bibinfo {volume} {12}},\ \bibinfo {pages}
		{7372} (\bibinfo {year} {2016}{\natexlab{b}})}\BibitemShut {NoStop}%
	\bibitem [{\citenamefont {Clop{\'e}s}\ \emph {et~al.}(2020)\citenamefont
		{Clop{\'e}s}, \citenamefont {Gompper},\ and\ \citenamefont
		{Winkler}}]{clop:20}%
	\BibitemOpen
	\bibfield  {author} {\bibinfo {author} {\bibfnamefont {J.}~\bibnamefont
			{Clop{\'e}s}}, \bibinfo {author} {\bibfnamefont {G.}~\bibnamefont
			{Gompper}},\ and\ \bibinfo {author} {\bibfnamefont {R.~G.}\ \bibnamefont
			{Winkler}},\ }\href {https://doi.org/10.1039/D0SM01569E} {\bibfield
		{journal} {\bibinfo  {journal} {Soft Matter}\ }\textbf {\bibinfo {volume}
			{16}},\ \bibinfo {pages} {10676} (\bibinfo {year} {2020})}\BibitemShut
	{NoStop}%
	\bibitem [{\citenamefont {Saintillan}\ and\ \citenamefont
		{Shelley}(2008)}]{sain:08}%
	\BibitemOpen
	\bibfield  {author} {\bibinfo {author} {\bibfnamefont {D.}~\bibnamefont
			{Saintillan}}\ and\ \bibinfo {author} {\bibfnamefont {M.~J.}\ \bibnamefont
			{Shelley}},\ }\href {https://doi.org/10.1103/PhysRevLett.100.178103}
	{\bibfield  {journal} {\bibinfo  {journal} {Phys. Rev. Lett.}\ }\textbf
		{\bibinfo {volume} {100}},\ \bibinfo {pages} {178103} (\bibinfo {year}
		{2008})}\BibitemShut {NoStop}%
	\bibitem [{\citenamefont {Turner}\ \emph {et~al.}(2000)\citenamefont {Turner},
		\citenamefont {Ryu},\ and\ \citenamefont {Berg}}]{turn:00}%
	\BibitemOpen
	\bibfield  {author} {\bibinfo {author} {\bibfnamefont {L.}~\bibnamefont
			{Turner}}, \bibinfo {author} {\bibfnamefont {W.~S.}\ \bibnamefont {Ryu}},\
		and\ \bibinfo {author} {\bibfnamefont {H.~C.}\ \bibnamefont {Berg}},\ }\href
	{https://doi.org/10.1128/jb.182.10.2793-2801.2000} {\bibfield  {journal}
		{\bibinfo  {journal} {J. Bacteriol.}\ }\textbf {\bibinfo {volume} {182}},\
		\bibinfo {pages} {2793} (\bibinfo {year} {2000})}\BibitemShut {NoStop}%
	\bibitem [{\citenamefont {Bianchi}\ \emph {et~al.}(2017)\citenamefont
		{Bianchi}, \citenamefont {Saglimbeni},\ and\ \citenamefont
		{Di~Leonardo}}]{bian:17}%
	\BibitemOpen
	\bibfield  {author} {\bibinfo {author} {\bibfnamefont {S.}~\bibnamefont
			{Bianchi}}, \bibinfo {author} {\bibfnamefont {F.}~\bibnamefont
			{Saglimbeni}},\ and\ \bibinfo {author} {\bibfnamefont {R.}~\bibnamefont
			{Di~Leonardo}},\ }\href {https://doi.org/10.1103/PhysRevX.7.011010}
	{\bibfield  {journal} {\bibinfo  {journal} {Phys. Rev. X}\ }\textbf {\bibinfo
			{volume} {7}},\ \bibinfo {pages} {011010} (\bibinfo {year}
		{2017})}\BibitemShut {NoStop}%
	\bibitem [{\citenamefont {Mathijssen}\ \emph {et~al.}(2019)\citenamefont
		{Mathijssen}, \citenamefont {Figueroa-Morales}, \citenamefont {Junot},
		\citenamefont {Cl\'ement}, \citenamefont {Lindner},\ and\ \citenamefont
		{Z{\"o}ttl}}]{math:19}%
	\BibitemOpen
	\bibfield  {author} {\bibinfo {author} {\bibfnamefont {A.~J. T.~M.}\
			\bibnamefont {Mathijssen}}, \bibinfo {author} {\bibfnamefont
			{N.}~\bibnamefont {Figueroa-Morales}}, \bibinfo {author} {\bibfnamefont
			{G.}~\bibnamefont {Junot}}, \bibinfo {author} {\bibfnamefont
			{E.}~\bibnamefont {Cl\'ement}}, \bibinfo {author} {\bibfnamefont
			{A.}~\bibnamefont {Lindner}},\ and\ \bibinfo {author} {\bibfnamefont
			{A.}~\bibnamefont {Z{\"o}ttl}},\ }\href
	{https://doi.org/10.1038/s41467-019-11360-0} {\bibfield  {journal} {\bibinfo
			{journal} {Nat. Commun.}\ }\textbf {\bibinfo {volume} {20}},\ \bibinfo
		{pages} {3434} (\bibinfo {year} {2019})}\BibitemShut {NoStop}%
	\bibitem [{\citenamefont {Mousavi}\ \emph {et~al.}(2020)\citenamefont
		{Mousavi}, \citenamefont {Gompper},\ and\ \citenamefont {Winkler}}]{mous:20}%
	\BibitemOpen
	\bibfield  {author} {\bibinfo {author} {\bibfnamefont {S.~M.}\ \bibnamefont
			{Mousavi}}, \bibinfo {author} {\bibfnamefont {G.}~\bibnamefont {Gompper}},\
		and\ \bibinfo {author} {\bibfnamefont {R.~G.}\ \bibnamefont {Winkler}},\
	}\href {https://doi.org/10.1039/D0SM00571A} {\bibfield  {journal} {\bibinfo
			{journal} {Soft Matter}\ }\textbf {\bibinfo {volume} {16}},\ \bibinfo {pages}
		{4866} (\bibinfo {year} {2020})}\BibitemShut {NoStop}%
	\bibitem [{\citenamefont {Clop{\'e}s}\ and\ \citenamefont
		{Winkler}(2021)}]{clop:21}%
	\BibitemOpen
	\bibfield  {author} {\bibinfo {author} {\bibfnamefont {J.}~\bibnamefont
			{Clop{\'e}s}}\ and\ \bibinfo {author} {\bibfnamefont {R.~G.}\ \bibnamefont
			{Winkler}},\ }\href {https://doi.org/10.1140/epje/s10189-021-00027-8}
	{\bibfield  {journal} {\bibinfo  {journal} {Eur. Phys. J. E}\ }\textbf
		{\bibinfo {volume} {44}},\ \bibinfo {pages} {17} (\bibinfo {year}
		{2021})}\BibitemShut {NoStop}%
	\bibitem [{\citenamefont {Cortese}\ and\ \citenamefont {Wan}(2021)}]{cort:21}%
	\BibitemOpen
	\bibfield  {author} {\bibinfo {author} {\bibfnamefont {D.}~\bibnamefont
			{Cortese}}\ and\ \bibinfo {author} {\bibfnamefont {K.~Y.}\ \bibnamefont
			{Wan}},\ }\href {https://doi.org/10.1103/PhysRevLett.126.088003} {\bibfield
		{journal} {\bibinfo  {journal} {Phys. Rev. Lett.}\ }\textbf {\bibinfo
			{volume} {126}},\ \bibinfo {pages} {088003} (\bibinfo {year}
		{2021})}\BibitemShut {NoStop}%
	\bibitem [{\citenamefont {Ballerini}\ \emph {et~al.}(2008)\citenamefont
		{Ballerini}, \citenamefont {Cabibbo}, \citenamefont {Candelier},
		\citenamefont {Cavagna}, \citenamefont {Cisbani}, \citenamefont {Giardina},
		\citenamefont {Orlandi}, \citenamefont {Parisi}, \citenamefont {Procaccini},
		\citenamefont {Viale},\ and\ \citenamefont {Zdravkovic}}]{ball:08.1}%
	\BibitemOpen
	\bibfield  {author} {\bibinfo {author} {\bibfnamefont {M.}~\bibnamefont
			{Ballerini}}, \bibinfo {author} {\bibfnamefont {N.}~\bibnamefont {Cabibbo}},
		\bibinfo {author} {\bibfnamefont {R.}~\bibnamefont {Candelier}}, \bibinfo
		{author} {\bibfnamefont {A.}~\bibnamefont {Cavagna}}, \bibinfo {author}
		{\bibfnamefont {E.}~\bibnamefont {Cisbani}}, \bibinfo {author} {\bibfnamefont
			{I.}~\bibnamefont {Giardina}}, \bibinfo {author} {\bibfnamefont
			{A.}~\bibnamefont {Orlandi}}, \bibinfo {author} {\bibfnamefont
			{G.}~\bibnamefont {Parisi}}, \bibinfo {author} {\bibfnamefont
			{A.}~\bibnamefont {Procaccini}}, \bibinfo {author} {\bibfnamefont
			{M.}~\bibnamefont {Viale}},\ and\ \bibinfo {author} {\bibfnamefont
			{V.}~\bibnamefont {Zdravkovic}},\ }\href
	{https://doi.org/10.1016/j.anbehav.2008.02.004} {\bibfield  {journal}
		{\bibinfo  {journal} {Anim. Behav.}\ }\textbf {\bibinfo {volume} {76}},\
		\bibinfo {pages} {201} (\bibinfo {year} {2008})}\BibitemShut {NoStop}%
	\bibitem [{\citenamefont {Cavagna}\ and\ \citenamefont
		{Giardina}(2014)}]{cava:14}%
	\BibitemOpen
	\bibfield  {author} {\bibinfo {author} {\bibfnamefont {A.}~\bibnamefont
			{Cavagna}}\ and\ \bibinfo {author} {\bibfnamefont {I.}~\bibnamefont
			{Giardina}},\ }\href
	{https://doi.org/10.1146/annurev-conmatphys-031113-133834} {\bibfield
		{journal} {\bibinfo  {journal} {Annu. Rev. Condens. Matter Phys.}\ }\textbf
		{\bibinfo {volume} {5}},\ \bibinfo {pages} {183} (\bibinfo {year}
		{2014})}\BibitemShut {NoStop}%
	\bibitem [{\citenamefont {Papadopoulou}\ \emph {et~al.}(2022)\citenamefont
		{Papadopoulou}, \citenamefont {Hildenbrandt}, \citenamefont {Sankey},
		\citenamefont {Portugal},\ and\ \citenamefont {Hemelrijk}}]{papa:22}%
	\BibitemOpen
	\bibfield  {author} {\bibinfo {author} {\bibfnamefont {M.}~\bibnamefont
			{Papadopoulou}}, \bibinfo {author} {\bibfnamefont {H.}~\bibnamefont
			{Hildenbrandt}}, \bibinfo {author} {\bibfnamefont {D.~W.~E.}\ \bibnamefont
			{Sankey}}, \bibinfo {author} {\bibfnamefont {S.~J.}\ \bibnamefont
			{Portugal}},\ and\ \bibinfo {author} {\bibfnamefont {C.~K.}\ \bibnamefont
			{Hemelrijk}},\ }\href {https://doi.org/10.1371/journal.pcbi.1009772}
	{\bibfield  {journal} {\bibinfo  {journal} {PLOS Computational Biology}\
		}\textbf {\bibinfo {volume} {18}},\ \bibinfo {pages} {e1009772} (\bibinfo
		{year} {2022})}\BibitemShut {NoStop}%
	\bibitem [{\citenamefont {Gompper}\ \emph {et~al.}(2020)\citenamefont
		{Gompper}, \citenamefont {Winkler}, \citenamefont {Speck}, \citenamefont
		{Solon}, \citenamefont {Nardini}, \citenamefont {Peruani}, \citenamefont
		{L{\"o}wen}, \citenamefont {Golestanian}, \citenamefont {Kaupp},
		\citenamefont {Alvarez}, \citenamefont {Ki{\o}rboe}, \citenamefont {Lauga},
		\citenamefont {Poon}, \citenamefont {DeSimone}, \citenamefont
		{Mui{\~n}os-Landin}, \citenamefont {Fischer}, \citenamefont {S{\"o}ker},
		\citenamefont {Cichos}, \citenamefont {Kapral}, \citenamefont {Gaspard},
		\citenamefont {Ripoll}, \citenamefont {Sagues}, \citenamefont
		{Doostmohammadi}, \citenamefont {Yeomans}, \citenamefont {Aranson},
		\citenamefont {Bechinger}, \citenamefont {Stark}, \citenamefont {Hemelrijk},
		\citenamefont {Nedelec}, \citenamefont {Sarkar}, \citenamefont {Aryaksama},
		\citenamefont {Lacroix}, \citenamefont {Duclos}, \citenamefont {Yashunsky},
		\citenamefont {Silberzan}, \citenamefont {Arroyo},\ and\ \citenamefont
		{Kale}}]{gomp:20}%
	\BibitemOpen
	\bibfield  {author} {\bibinfo {author} {\bibfnamefont {G.}~\bibnamefont
			{Gompper}}, \bibinfo {author} {\bibfnamefont {R.~G.}\ \bibnamefont
			{Winkler}}, \bibinfo {author} {\bibfnamefont {T.}~\bibnamefont {Speck}},
		\bibinfo {author} {\bibfnamefont {A.}~\bibnamefont {Solon}}, \bibinfo
		{author} {\bibfnamefont {C.}~\bibnamefont {Nardini}}, \bibinfo {author}
		{\bibfnamefont {F.}~\bibnamefont {Peruani}}, \bibinfo {author} {\bibfnamefont
			{H.}~\bibnamefont {L{\"o}wen}}, \bibinfo {author} {\bibfnamefont
			{R.}~\bibnamefont {Golestanian}}, \bibinfo {author} {\bibfnamefont {U.~B.}\
			\bibnamefont {Kaupp}}, \bibinfo {author} {\bibfnamefont {L.}~\bibnamefont
			{Alvarez}}, \bibinfo {author} {\bibfnamefont {T.}~\bibnamefont {Ki{\o}rboe}},
		\bibinfo {author} {\bibfnamefont {E.}~\bibnamefont {Lauga}}, \bibinfo
		{author} {\bibfnamefont {W.~C.~K.}\ \bibnamefont {Poon}}, \bibinfo {author}
		{\bibfnamefont {A.}~\bibnamefont {DeSimone}}, \bibinfo {author}
		{\bibfnamefont {S.}~\bibnamefont {Mui{\~n}os-Landin}}, \bibinfo {author}
		{\bibfnamefont {A.}~\bibnamefont {Fischer}}, \bibinfo {author} {\bibfnamefont
			{N.~A.}\ \bibnamefont {S{\"o}ker}}, \bibinfo {author} {\bibfnamefont
			{F.}~\bibnamefont {Cichos}}, \bibinfo {author} {\bibfnamefont
			{R.}~\bibnamefont {Kapral}}, \bibinfo {author} {\bibfnamefont
			{P.}~\bibnamefont {Gaspard}}, \bibinfo {author} {\bibfnamefont
			{M.}~\bibnamefont {Ripoll}}, \bibinfo {author} {\bibfnamefont
			{F.}~\bibnamefont {Sagues}}, \bibinfo {author} {\bibfnamefont
			{A.}~\bibnamefont {Doostmohammadi}}, \bibinfo {author} {\bibfnamefont
			{J.~M.}\ \bibnamefont {Yeomans}}, \bibinfo {author} {\bibfnamefont {I.~S.}\
			\bibnamefont {Aranson}}, \bibinfo {author} {\bibfnamefont {C.}~\bibnamefont
			{Bechinger}}, \bibinfo {author} {\bibfnamefont {H.}~\bibnamefont {Stark}},
		\bibinfo {author} {\bibfnamefont {C.~K.}\ \bibnamefont {Hemelrijk}}, \bibinfo
		{author} {\bibfnamefont {F.~J.}\ \bibnamefont {Nedelec}}, \bibinfo {author}
		{\bibfnamefont {T.}~\bibnamefont {Sarkar}}, \bibinfo {author} {\bibfnamefont
			{T.}~\bibnamefont {Aryaksama}}, \bibinfo {author} {\bibfnamefont
			{M.}~\bibnamefont {Lacroix}}, \bibinfo {author} {\bibfnamefont
			{G.}~\bibnamefont {Duclos}}, \bibinfo {author} {\bibfnamefont
			{V.}~\bibnamefont {Yashunsky}}, \bibinfo {author} {\bibfnamefont
			{P.}~\bibnamefont {Silberzan}}, \bibinfo {author} {\bibfnamefont
			{M.}~\bibnamefont {Arroyo}},\ and\ \bibinfo {author} {\bibfnamefont
			{S.}~\bibnamefont {Kale}},\ }\href {https://doi.org/10.1088/1361-648x/ab6348}
	{\bibfield  {journal} {\bibinfo  {journal} {J. Phys: Condens. Matter}\
		}\textbf {\bibinfo {volume} {32}},\ \bibinfo {pages} {193001} (\bibinfo
		{year} {2020})}\BibitemShut {NoStop}%
	\bibitem [{\citenamefont {Katz}\ \emph {et~al.}(2011)\citenamefont {Katz},
		\citenamefont {Tunstr{\o}m}, \citenamefont {Ioannou}, \citenamefont {Huepe},\
		and\ \citenamefont {Couzin}}]{katz:11}%
	\BibitemOpen
	\bibfield  {author} {\bibinfo {author} {\bibfnamefont {Y.}~\bibnamefont
			{Katz}}, \bibinfo {author} {\bibfnamefont {K.}~\bibnamefont {Tunstr{\o}m}},
		\bibinfo {author} {\bibfnamefont {C.~C.}\ \bibnamefont {Ioannou}}, \bibinfo
		{author} {\bibfnamefont {C.}~\bibnamefont {Huepe}},\ and\ \bibinfo {author}
		{\bibfnamefont {I.~D.}\ \bibnamefont {Couzin}},\ }\href
	{https://doi.org/10.1073/pnas.1107583108} {\bibfield  {journal} {\bibinfo
			{journal} {Proc. Natl. Acad. Sci. USA}\ }\textbf {\bibinfo {volume} {108}},\
		\bibinfo {pages} {18720} (\bibinfo {year} {2011})}\BibitemShut {NoStop}%
	\bibitem [{\citenamefont {Vicsek}\ and\ \citenamefont
		{Zafeiris}(2012)}]{vics:12}%
	\BibitemOpen
	\bibfield  {author} {\bibinfo {author} {\bibfnamefont {T.}~\bibnamefont
			{Vicsek}}\ and\ \bibinfo {author} {\bibfnamefont {A.}~\bibnamefont
			{Zafeiris}},\ }\href {https://doi.org/10.1016/j.physrep.2012.03.00}
	{\bibfield  {journal} {\bibinfo  {journal} {Phys. Rep.}\ }\textbf {\bibinfo
			{volume} {517}},\ \bibinfo {pages} {71} (\bibinfo {year} {2012})}\BibitemShut
	{NoStop}%
	\bibitem [{\citenamefont {Toner}\ and\ \citenamefont {Tu}(1995)}]{tone:95}%
	\BibitemOpen
	\bibfield  {author} {\bibinfo {author} {\bibfnamefont {J.}~\bibnamefont
			{Toner}}\ and\ \bibinfo {author} {\bibfnamefont {Y.}~\bibnamefont {Tu}},\
	}\href {https://doi.org/10.1103/PhysRevLett.75.4326} {\bibfield  {journal}
		{\bibinfo  {journal} {Phys. Rev. Lett.}\ }\textbf {\bibinfo {volume} {75}},\
		\bibinfo {pages} {4326} (\bibinfo {year} {1995})}\BibitemShut {NoStop}%
	\bibitem [{\citenamefont {Aditi~Simha}\ and\ \citenamefont
		{Ramaswamy}(2002)}]{simh:02}%
	\BibitemOpen
	\bibfield  {author} {\bibinfo {author} {\bibfnamefont {R.}~\bibnamefont
			{Aditi~Simha}}\ and\ \bibinfo {author} {\bibfnamefont {S.}~\bibnamefont
			{Ramaswamy}},\ }\href {https://doi.org/10.1103/PhysRevLett.89.058101}
	{\bibfield  {journal} {\bibinfo  {journal} {Phys. Rev. Lett.}\ }\textbf
		{\bibinfo {volume} {89}},\ \bibinfo {pages} {058101} (\bibinfo {year}
		{2002})}\BibitemShut {NoStop}%
	\bibitem [{\citenamefont {Ekiel-Je{\.z}ewska}\ \emph
		{et~al.}(2008)\citenamefont {Ekiel-Je{\.z}ewska}, \citenamefont {Gubiec},\
		and\ \citenamefont {Szymczak}}]{ekie:08}%
	\BibitemOpen
	\bibfield  {author} {\bibinfo {author} {\bibfnamefont {M.~L.}\ \bibnamefont
			{Ekiel-Je{\.z}ewska}}, \bibinfo {author} {\bibfnamefont {T.}~\bibnamefont
			{Gubiec}},\ and\ \bibinfo {author} {\bibfnamefont {P.}~\bibnamefont
			{Szymczak}},\ }\href {https://doi.org/10.1063/1.2930881} {\bibfield
		{journal} {\bibinfo  {journal} {Phys. Fluids}\ }\textbf {\bibinfo {volume}
			{20}},\ \bibinfo {pages} {063102} (\bibinfo {year} {2008})}\BibitemShut
	{NoStop}%
	\bibitem [{\citenamefont {Saggiorato}\ \emph {et~al.}(2015)\citenamefont
		{Saggiorato}, \citenamefont {Elgeti}, \citenamefont {Winkler},\ and\
		\citenamefont {Gompper}}]{sagg:15}%
	\BibitemOpen
	\bibfield  {author} {\bibinfo {author} {\bibfnamefont {G.}~\bibnamefont
			{Saggiorato}}, \bibinfo {author} {\bibfnamefont {J.}~\bibnamefont {Elgeti}},
		\bibinfo {author} {\bibfnamefont {R.~G.}\ \bibnamefont {Winkler}},\ and\
		\bibinfo {author} {\bibfnamefont {G.}~\bibnamefont {Gompper}},\ }\href
	{https://doi.org/10.1039/C5SM01069A} {\bibfield  {journal} {\bibinfo
			{journal} {Soft Matter}\ }\textbf {\bibinfo {volume} {11}},\ \bibinfo {pages}
		{7337} (\bibinfo {year} {2015})}\BibitemShut {NoStop}%
	\bibitem [{\citenamefont {Qi}\ \emph {et~al.}(2020)\citenamefont {Qi},
		\citenamefont {Westphal}, \citenamefont {Gompper},\ and\ \citenamefont
		{Winkler}}]{qi:20}%
	\BibitemOpen
	\bibfield  {author} {\bibinfo {author} {\bibfnamefont {K.}~\bibnamefont
			{Qi}}, \bibinfo {author} {\bibfnamefont {E.}~\bibnamefont {Westphal}},
		\bibinfo {author} {\bibfnamefont {G.}~\bibnamefont {Gompper}},\ and\ \bibinfo
		{author} {\bibfnamefont {R.~G.}\ \bibnamefont {Winkler}},\ }\href
	{https://doi.org/10.1103/PhysRevLett.124.068001} {\bibfield  {journal}
		{\bibinfo  {journal} {Phys. Rev. Lett.}\ }\textbf {\bibinfo {volume} {124}},\
		\bibinfo {pages} {068001} (\bibinfo {year} {2020})}\BibitemShut {NoStop}%
	\bibitem [{\citenamefont {Goldstein}\ \emph {et~al.}(2009)\citenamefont
		{Goldstein}, \citenamefont {Polin},\ and\ \citenamefont {Tuval}}]{gold:09}%
	\BibitemOpen
	\bibfield  {author} {\bibinfo {author} {\bibfnamefont {R.~E.}\ \bibnamefont
			{Goldstein}}, \bibinfo {author} {\bibfnamefont {M.}~\bibnamefont {Polin}},\
		and\ \bibinfo {author} {\bibfnamefont {I.}~\bibnamefont {Tuval}},\ }\href
	{https://doi.org/10.1103/PhysRevLett.103.168103} {\bibfield  {journal}
		{\bibinfo  {journal} {Phys. Rev. Lett.}\ }\textbf {\bibinfo {volume} {103}},\
		\bibinfo {pages} {168103} (\bibinfo {year} {2009})}\BibitemShut {NoStop}%
	\bibitem [{\citenamefont {Reigh}\ \emph {et~al.}(2012)\citenamefont {Reigh},
		\citenamefont {Winkler},\ and\ \citenamefont {Gompper}}]{reig:12}%
	\BibitemOpen
	\bibfield  {author} {\bibinfo {author} {\bibfnamefont {S.~Y.}\ \bibnamefont
			{Reigh}}, \bibinfo {author} {\bibfnamefont {R.~G.}\ \bibnamefont {Winkler}},\
		and\ \bibinfo {author} {\bibfnamefont {G.}~\bibnamefont {Gompper}},\ }\href
	{https://doi.org/10.1039/C2SM07378A} {\bibfield  {journal} {\bibinfo
			{journal} {Soft Matter}\ }\textbf {\bibinfo {volume} {8}},\ \bibinfo {pages}
		{4363} (\bibinfo {year} {2012})}\BibitemShut {NoStop}%
	\bibitem [{\citenamefont {Geyer}\ \emph {et~al.}(2013)\citenamefont {Geyer},
		\citenamefont {J{\"u}licher}, \citenamefont {Howard},\ and\ \citenamefont
		{Friedrich}}]{geye:13}%
	\BibitemOpen
	\bibfield  {author} {\bibinfo {author} {\bibfnamefont {V.~F.}\ \bibnamefont
			{Geyer}}, \bibinfo {author} {\bibfnamefont {F.}~\bibnamefont {J{\"u}licher}},
		\bibinfo {author} {\bibfnamefont {J.}~\bibnamefont {Howard}},\ and\ \bibinfo
		{author} {\bibfnamefont {B.~M.}\ \bibnamefont {Friedrich}},\ }\href
	{https://doi.org/10.1073/pnas.1300895110} {\bibfield  {journal} {\bibinfo
			{journal} {Proc. Natl. Acad. Sci. USA}\ }\textbf {\bibinfo {volume} {110}},\
		\bibinfo {pages} {18058} (\bibinfo {year} {2013})}\BibitemShut {NoStop}%
	\bibitem [{\citenamefont {Brumley}\ \emph {et~al.}(2014)\citenamefont
		{Brumley}, \citenamefont {Wan}, \citenamefont {Polin},\ and\ \citenamefont
		{Goldstein}}]{brum:14}%
	\BibitemOpen
	\bibfield  {author} {\bibinfo {author} {\bibfnamefont {D.~R.}\ \bibnamefont
			{Brumley}}, \bibinfo {author} {\bibfnamefont {K.~Y.}\ \bibnamefont {Wan}},
		\bibinfo {author} {\bibfnamefont {M.}~\bibnamefont {Polin}},\ and\ \bibinfo
		{author} {\bibfnamefont {R.~E.}\ \bibnamefont {Goldstein}},\ }\href
	{https://doi.org/10.7554/eLife.02750} {\bibfield  {journal} {\bibinfo
			{journal} {eLife}\ }\textbf {\bibinfo {volume} {3}},\ \bibinfo {pages}
		{e02750} (\bibinfo {year} {2014})}\BibitemShut {NoStop}%
	\bibitem [{\citenamefont {Theers}\ and\ \citenamefont
		{Winkler}(2014)}]{thee:14}%
	\BibitemOpen
	\bibfield  {author} {\bibinfo {author} {\bibfnamefont {M.}~\bibnamefont
			{Theers}}\ and\ \bibinfo {author} {\bibfnamefont {R.~G.}\ \bibnamefont
			{Winkler}},\ }\href {https://doi.org/10.1039/C4SM00770K} {\bibfield
		{journal} {\bibinfo  {journal} {Soft Matter}\ }\textbf {\bibinfo {volume}
			{10}},\ \bibinfo {pages} {5894} (\bibinfo {year} {2014})}\BibitemShut
	{NoStop}%
	\bibitem [{\citenamefont {Eisenstecken}\ \emph {et~al.}(2016)\citenamefont
		{Eisenstecken}, \citenamefont {Gompper},\ and\ \citenamefont
		{Winkler}}]{eise:16}%
	\BibitemOpen
	\bibfield  {author} {\bibinfo {author} {\bibfnamefont {T.}~\bibnamefont
			{Eisenstecken}}, \bibinfo {author} {\bibfnamefont {G.}~\bibnamefont
			{Gompper}},\ and\ \bibinfo {author} {\bibfnamefont {R.~G.}\ \bibnamefont
			{Winkler}},\ }\href {https://doi.org/10.3390/polym8080304} {\bibfield
		{journal} {\bibinfo  {journal} {Polymers}\ }\textbf {\bibinfo {volume} {8}},\
		\bibinfo {pages} {304} (\bibinfo {year} {2016})}\BibitemShut {NoStop}%
	\bibitem [{\citenamefont {Hu}\ \emph {et~al.}(2015{\natexlab{b}})\citenamefont
		{Hu}, \citenamefont {Wysocki}, \citenamefont {Winkler},\ and\ \citenamefont
		{Gompper}}]{hu:15}%
	\BibitemOpen
	\bibfield  {author} {\bibinfo {author} {\bibfnamefont {J.}~\bibnamefont
			{Hu}}, \bibinfo {author} {\bibfnamefont {A.}~\bibnamefont {Wysocki}},
		\bibinfo {author} {\bibfnamefont {R.~G.}\ \bibnamefont {Winkler}},\ and\
		\bibinfo {author} {\bibfnamefont {G.}~\bibnamefont {Gompper}},\ }\href
	{https://doi.org/10.1038/srep09586} {\bibfield  {journal} {\bibinfo
			{journal} {Sci. Rep.}\ }\textbf {\bibinfo {volume} {5}},\ \bibinfo {pages}
		{9586} (\bibinfo {year} {2015}{\natexlab{b}})}\BibitemShut {NoStop}%
	\bibitem [{\citenamefont {Babu}\ and\ \citenamefont {Stark}(2012)}]{babu:12}%
	\BibitemOpen
	\bibfield  {author} {\bibinfo {author} {\bibfnamefont {S.~B.}\ \bibnamefont
			{Babu}}\ and\ \bibinfo {author} {\bibfnamefont {H.}~\bibnamefont {Stark}},\
	}\href {https://doi.org/10.1088/1367-2630/14/8/085012} {\bibfield  {journal}
		{\bibinfo  {journal} {New J. Phys.}\ }\textbf {\bibinfo {volume} {14}},\
		\bibinfo {pages} {085012} (\bibinfo {year} {2012})}\BibitemShut {NoStop}%
	\bibitem [{\citenamefont {Rode}\ \emph {et~al.}(2019)\citenamefont {Rode},
		\citenamefont {Elgeti},\ and\ \citenamefont {Gompper}}]{rode:19}%
	\BibitemOpen
	\bibfield  {author} {\bibinfo {author} {\bibfnamefont {S.}~\bibnamefont
			{Rode}}, \bibinfo {author} {\bibfnamefont {J.}~\bibnamefont {Elgeti}},\ and\
		\bibinfo {author} {\bibfnamefont {G.}~\bibnamefont {Gompper}},\ }\href
	{https://doi.org/10.1088/1367-2630/aaf544} {\bibfield  {journal} {\bibinfo
			{journal} {New J. Phys.}\ }\textbf {\bibinfo {volume} {21}},\ \bibinfo
		{pages} {013016} (\bibinfo {year} {2019})}\BibitemShut {NoStop}%
	\bibitem [{\citenamefont {Noguchi}\ and\ \citenamefont
		{Gompper}(2008)}]{nogu:08}%
	\BibitemOpen
	\bibfield  {author} {\bibinfo {author} {\bibfnamefont {H.}~\bibnamefont
			{Noguchi}}\ and\ \bibinfo {author} {\bibfnamefont {G.}~\bibnamefont
			{Gompper}},\ }\href@noop {} {\bibfield  {journal} {\bibinfo  {journal} {Phys.
				Rev. E}\ }\textbf {\bibinfo {volume} {78}},\ \bibinfo {pages} {016706}
		(\bibinfo {year} {2008})}\BibitemShut {NoStop}%
	\bibitem [{\citenamefont {Ihle}\ and\ \citenamefont {Kroll}(2003)}]{ihle:03}%
	\BibitemOpen
	\bibfield  {author} {\bibinfo {author} {\bibfnamefont {T.}~\bibnamefont
			{Ihle}}\ and\ \bibinfo {author} {\bibfnamefont {D.~M.}\ \bibnamefont
			{Kroll}},\ }\href@noop {} {\bibfield  {journal} {\bibinfo  {journal} {Phys.
				Rev. E}\ }\textbf {\bibinfo {volume} {67}},\ \bibinfo {pages} {066705}
		(\bibinfo {year} {2003})}\BibitemShut {NoStop}%
	\bibitem [{\citenamefont {Huang}\ \emph {et~al.}(2010)\citenamefont {Huang},
		\citenamefont {Chatterji}, \citenamefont {Sutmann}, \citenamefont {Gompper},\
		and\ \citenamefont {Winkler}}]{huan:10.1}%
	\BibitemOpen
	\bibfield  {author} {\bibinfo {author} {\bibfnamefont {C.-C.}\ \bibnamefont
			{Huang}}, \bibinfo {author} {\bibfnamefont {A.}~\bibnamefont {Chatterji}},
		\bibinfo {author} {\bibfnamefont {G.}~\bibnamefont {Sutmann}}, \bibinfo
		{author} {\bibfnamefont {G.}~\bibnamefont {Gompper}},\ and\ \bibinfo {author}
		{\bibfnamefont {R.~G.}\ \bibnamefont {Winkler}},\ }\href
	{https://doi.org/10.1016/j.jcp.2009.09.024} {\bibfield  {journal} {\bibinfo
			{journal} {J. Comput. Phys.}\ }\textbf {\bibinfo {volume} {229}},\ \bibinfo
		{pages} {168} (\bibinfo {year} {2010})}\BibitemShut {NoStop}%
	\bibitem [{\citenamefont {Westphal}\ \emph {et~al.}(2014)\citenamefont
		{Westphal}, \citenamefont {Singh}, \citenamefont {Huang}, \citenamefont
		{Gompper},\ and\ \citenamefont {Winkler}}]{west:14}%
	\BibitemOpen
	\bibfield  {author} {\bibinfo {author} {\bibfnamefont {E.}~\bibnamefont
			{Westphal}}, \bibinfo {author} {\bibfnamefont {S.~P.}\ \bibnamefont {Singh}},
		\bibinfo {author} {\bibfnamefont {C.-C.}\ \bibnamefont {Huang}}, \bibinfo
		{author} {\bibfnamefont {G.}~\bibnamefont {Gompper}},\ and\ \bibinfo {author}
		{\bibfnamefont {R.~G.}\ \bibnamefont {Winkler}},\ }\href
	{https://doi.org/10.1016/j.cpc.2013.10.004} {\bibfield  {journal} {\bibinfo
			{journal} {Comput. Phys. Comm.}\ }\textbf {\bibinfo {volume} {185}},\
		\bibinfo {pages} {495} (\bibinfo {year} {2014})}\BibitemShut {NoStop}%
	\bibitem [{\citenamefont {Theers}\ and\ \citenamefont
		{Winkler}(2015)}]{thee:15}%
	\BibitemOpen
	\bibfield  {author} {\bibinfo {author} {\bibfnamefont {M.}~\bibnamefont
			{Theers}}\ and\ \bibinfo {author} {\bibfnamefont {R.~G.}\ \bibnamefont
			{Winkler}},\ }\href {https://doi.org/10.1103/PhysRevE.91.033309} {\bibfield
		{journal} {\bibinfo  {journal} {Phys. Rev. E}\ }\textbf {\bibinfo {volume}
			{91}},\ \bibinfo {pages} {033309} (\bibinfo {year} {2015})}\BibitemShut
	{NoStop}%
\end{thebibliography}

%

\end{document}